\documentclass[english,showfigures,singlespace]{generic_preprint}

\usepackage{graphicx}
\usepackage{natbib}
\usepackage[colorlinks=true,linkcolor=blue,citecolor=blue,linktoc=none,pdftitle={LFP beta amplitude is predictive of mesoscopic spatio-temporal phase patterns},pdfauthor={Denker et al.}]{hyperref}
\usepackage{varioref}

\usepackage[colorinlistoftodos,color=green!20,textsize=small,prependcaption={},textwidth=4cm]{todonotes}
\presetkeys{todonotes}{inline,backgroundcolor=yellow}{}
    
\usepackage{refstyle}
\let\eqref=\relax
\newref{sec}{Name=Section~,Names=Sections~,name=Section~,names=Sections~}
\newref{sub}{Name=Section~,Names=Sections~,name=Section~,names=Sections~}
\newref{eq}{Name=Eq.~,Names=Eqs.~,name=eq.~,names=eqs.~}
\newref{fig}{Name=Figure~,Names=Figures~,name=Figure~,names=Figures~}
\newref{tab}{Name=Table~,Names=Tables~,name=Table~,names=Tables~}

\newcommand{\myunits}[2]{$#1\ \mathrm{#2}$}

\newcommand{\ram}[1]{A_{#1}(t)}
\newcommand{\rpm}[1]{\Phi_{#1}(t)}
\newcommand{\pgm}[1]{\Gamma_{#1}(t)}
\newcommand{\pvm}[1]{\Psi_{#1}(t)}
\newcommand{\pdm}[1]{\Delta_{#1}(t)}
\newcommand{\gcm}[1]{\Lambda_{#1}(t)}

\newcommand{\cvp}{\sigma_p(t)}
\newcommand{\cvg}{\sigma_g(t)}
\newcommand{\mlc}{\mu_c(t)}
\newcommand{\cnt}{C(t)}
\newcommand{\prl}{R_\parallel}
\newcommand{\ort}{R_\perp}

\newcommand{\velocity}{v(t)}
\newcommand{\amplitude}{a(t)}
\newcommand{\direction}{d(t)}

\newcommand{\comm}[1]{}

\definecolor{mediumtealblue}{rgb}{0.0, 0.33, 0.71}
\definecolor{debianred}{rgb}{0.84, 0.04, 0.33}
\usepackage{soul}

\begin{document}

%
%

\title{LFP beta amplitude is predictive of mesoscopic spatio-temporal phase patterns}

\author{Michael Denker\textsuperscript{1},
Lyuba Zehl\textsuperscript{1},
Bj\o rg E. Kilavik\textsuperscript{2},
Markus Diesmann\textsuperscript{1},
Thomas Brochier\textsuperscript{2},
Alexa Riehle\textsuperscript{2,1,3}, and
Sonja Gr\"un\textsuperscript{1,3,4}}

\maketitle

\begin{affiliations}
\item Institute of Neuroscience and Medicine (INM-6) and Institute for Advanced Simulation (IAS-6) and
\item[] JARA BRAIN Institute I, J\"ulich Research Centre, J\"ulich, Germany
\item Institut de Neurosciences de la Timone (INT), CNRS-Aix-Marseille University, UMR 7289, Marseille, France
\item RIKEN Brain Science Institute, Wako City, Japan
\item Theoretical Systems Neurobiology, RWTH Aachen University, Germany
\end{affiliations}

%
%
\begin{abstract}
Beta oscillations observed in motor cortical local field potentials (LFPs) recorded on separate electrodes of a multi-electrode array have been shown to exhibit non-zero phase shifts that organize into a planar wave propagation. Here, we generalize this concept by introducing additional classes of patterns that fully describe the spatial organization of beta oscillations. During a delayed reach-to-grasp task in monkey primary motor and dorsal premotor cortices we distinguish planar, synchronized, random, circular, and radial phase patterns. We observe that specific patterns correlate with the beta amplitude (envelope). In particular, wave propagation accelerates with growing amplitude, and culminates at maximum amplitude in a synchronized pattern. Furthermore, the occurrence probability of a particular pattern is modulated with behavioral epochs: Planar waves and synchronized patterns are more present during  movement preparation where beta amplitudes are large, whereas random phase patterns are dominant during  movement execution where beta amplitudes are small.
\end{abstract}

%
%
\begin{introduction}
\label{sec:Introduction}
\pdfbookmark[0]{\introtitle}{sec:Introduction-pdf}

The local field potential (LFP) has long served as a readily available brain signal to monitor the average input activity that reaches the neurons in the vicinity of extracellular recording electrodes \citep{Mitzdorf85_37,Logothetis04_735,Einevoll13_770}. A hallmark of the LFP is the ubiquitous presence of oscillations in various frequency bands \citep{Buzsaki04_1926} modulating in time and across different brain structures. These oscillations have been linked to a variety of brain processes such as attention \citep{Fries01_1560}, stimulus encoding \citep{Engel90_588}, or memory formation \citep{Pesaran02_805,Dotson14_13600}. These findings support the basis of modern theories concerning the functional implication of oscillatory brain activities, such as feature binding \citep{Singer99_49}, the concept of communication-through-coherence \citep{Fries05_474,Fries15_220,Womelsdorf07_1609}, the phase-of-firing coding \citep{Masquelier09_13484}, or predictive coding \citep{Friston15_1}. In motor cortex, beta oscillations (in the range of \myunits{15-35}{Hz}) are one of the most prominent types of oscillatory activity. They have been linked to states of general arousal, movement preparation, or postural maintenance (\citealt{Kilavik12_2148}; review in \citealt{Kilavik13_15}), and are typically suppressed during active movement (cf.~\citealt{Pfurtscheller79_138,Rougeul79_310}).

Technological progress recently led to the development of multi-electrode arrays enabling neuroscientists to record massively parallel neuronal signals in a precisely identifiable spatial arrangement. Although LFP signals recorded in motor cortex from electrodes separated by up to several millimeters are typically highly correlated \citep{Murthy96_3949}, the analysis of the instantaneous phase of the oscillation \citep[e.g.,][]{Varela01_229} revealed a non-zero temporal shift between electrodes \citep{Denker11_2681}. Such shifts may be expressed by the formation of dynamic spatial patterns propagating along preferred directions across the cortical surface, referred to as traveling waves \citep{Rubino06_1549}. Indeed, the phenomenon of traveling waves has been described in multiple brain areas, such as the visual cortex (\citealt{Nauhaus09_70,Zanos15_615}; see for a review \citealt{Sato12_218}), the olfactory bulb \citep{Freeman78_586,Friedrich04_862}, or the thalamus \citep{Kim95_1301}. However, the type of wave activity observed in motor cortex differs from the types of traveling waves described in visual cortex, for instance, by using optical imaging techniques \citep{Muller14_3675}. In this latter study the authors described a single-cycle propagation of elevated activity from a central hotspot outwards which was either induced by stimulation or occurred spontaneously. In contrast, motor cortical waves were described so-far as rather being unidirectional throughout the cortical region covered by \myunits{4\textrm{-by-}4}{mm} multi-electrode arrays. These waves traveling homogeneously along a defined direction are generally called planar waves. The probability of observing these planar waves may rapidly change as a function of behavioral context. Indeed, \citet{Rubino06_1549} found that the average coherence of phase gradients across electrodes, considered as being a signature of planar wave propagation, was highest at the beginning of the instructed delay of a center-out reaching task.

The planar waves described in \citet{Rubino06_1549} represent the most salient type of dynamic pattern formation, and are easily identifiable by the parallel arrangement of the phase gradients. However, potentially different patterns of spatial organization of beta oscillations outside periods of planar waves have not yet been described. It is reasonable to assume that oscillatory activities do exhibit other types of patterns commonly associated with theoretical systems displaying pattern formation \citep[e.g.,][]{Ermentrout01_33,Heitmann12_67}, such as divergences or singularities. In visual cortical area MT of the anesthetized marmoset monkey, for instance, \citet{Townsend15_4657} described a variety of such patterns in slow (delta) oscillations.

The occurrences of motor cortical planar waves seem to be of very short duration, in the order of \myunits{50}{ms}, as noted by \citet{Rubino06_1549} in their Supplemental Information. This is evocative to the finding that motor cortical beta oscillations strongly modulate their amplitude by exhibiting short-lasting high amplitude epochs of a few oscillatory cycles, the so-called spindles \citep{Murthy92_5670,Murthy96_3949}. Even though an individual beta spindle lasts far longer than the occurrence of a planar wave, their dynamic properties suggest that they are correlated. This hypothesis is further supported by the finding that when considering data of different trials, both traveling waves \citep{Rubino06_1549} and beta power \citep{Kilavik13_15} are most prominent during  an instructed delay of a motor task. Moreover, for slow oscillations, the power was found to correlate with the dynamics of activity patterns \citep{Townsend15_4657}.

The present work had three main goals: The first goal was to explore the possible presence of  wave-like spatio-temporal patterns other than planar waves. The second goal was to relate patterns to behavioral epochs in order to test their possible functional implication. The third goal was to test whether or not patterns were related to modulations in beta amplitude, both in single trials and across trials. Neuronal activity was recorded by using a 100-electrode Utah array, chronically implanted in primary motor (M1) and premotor (PM) cortices. Three monkeys were trained in an instructed-delay reach-to-grasp task \citep{Riehle13_48,Milekovic15_338}. We analyzed the spectral properties of the LFP signals and characterized the emergent spatio-temporal patterns based on the phase information. This analysis revealed a variety of spatio-temporal patterns in LFP beta oscillations that can  be clearly distinguished and identified as five categories of phase patterns. We developed statistical measures to identify the different phase patterns and the periods in which each of the patterns occurred, and determined their prevalence as a function of trial progression and behavioral epochs. Using these findings, we were able to establish the tight link between the modulation of LFP beta amplitude and the formation of spatio-temporal patterns of the oscillation. Preliminary results were presented in abstract form in \citep{Denker14_areadne,Denker15_sfn}.
\end{introduction}

%
%
\begin{results}
\label{sec:Results}
\pdfbookmark[0]{\resultstitle}{sec:Results-pdf}

Three monkeys were trained in a delayed reach-to-grasp task (\figref{spectral-properties}A) in which the animal had to grasp, pull and hold an object using either a side grip (SG) or a precision grip (PG), and either with a low force (LF) or high force (HF), resulting in a total of 4 pseudo-randomly presented trial types. The monkey was first presented with a cue for \myunits{300}{ms} which provided partial prior information either about the grip type (SG or PG) in grip-first trials, or the amount of force (LF or HF) in force-first trials, to be used in the upcoming movement. The cue was followed by a preparatory delay of \myunits{1}{s}. The GO signal, presented at the end of the delay, provided the missing information about either the force (LF or HF) or the grip type (SG or PG) in grip-first and force-first trials, respectively. The GO signal also instructed the monkey to initiate the reach and grasp movement.  Each correct trial was rewarded by a drop of apple sauce. \figref{spectral-properties}A shows the time line of the behavioral trial. The monkeys performed sessions of about \myunits{15}{min} (120--140 correct trials) in which either grip-first trials or force-first trials were requested. For a complete description of the behavioral task refer to the \nameref{sec:Methods}.

While the monkeys performed the task, neuronal activity was recorded using a 100-electrode Utah array (Blackrock Microsystems, Salt Lake City, UT, USA) implanted in the contralateral primary motor (MI) and premotor (PM) cortices with respect to the active arm (monkey L and N, left hemisphere, and monkey T, right hemisphere). The precise locations of the implanted arrays are shown in \figref{spectral-properties}B and C. In this study we concentrate on the local field potential (LFP) signals, filtered between \myunits{0.3 - 250}{Hz} and sampled at \myunits{1}{kHz}. We selected for further analysis in each monkey 15 recording sessions from the grip-first condition, and additionally 15 sessions from the force-first condition in monkey L and T, respectively. In the following, we start by characterizing the spectral properties of the recorded LFP activity to identify its oscillatory features, before quantifying these oscillations also in the spatial domain.

%
%
\begin{figure}
\centering
  \begin{minipage}[t]{0.63\textwidth}
    \ \\
    \includegraphics[width=\textwidth]{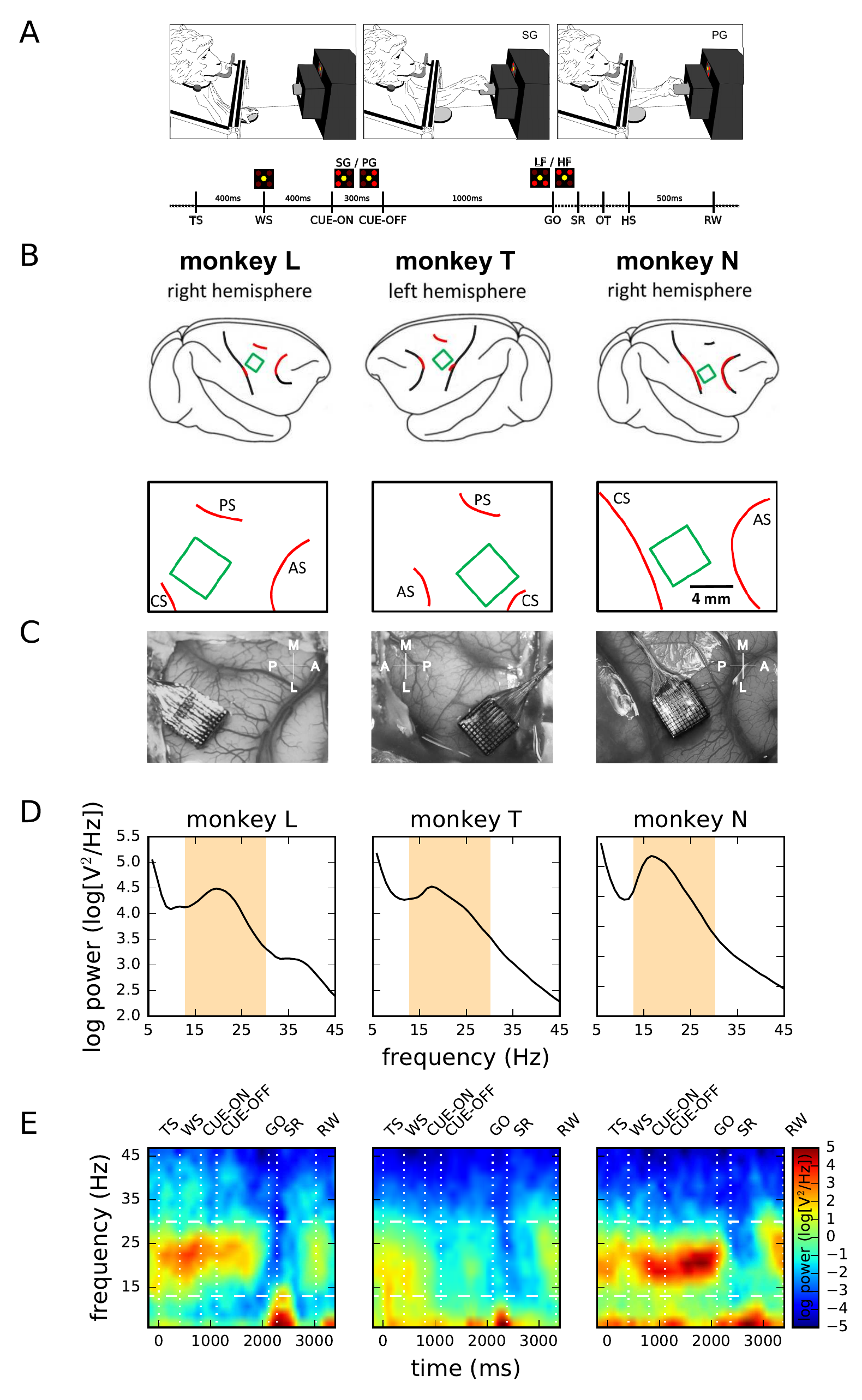}
  \end{minipage}\hfill
  \begin{minipage}[t]{0.34\textwidth}
    \caption{\textbf{Experimental task, array positions, and spectral properties of the LFP.} \textbf{A.} Task design. Top: sketch of the monkey during the task in the anticipatory position before GO (left), and while performing a side grip (middle) and precision grip (right).  Bottom: time line of the task. Labels indicate events (TS: trial start; WS: warning signal; CUE-ON/OFF: cue on/cue off; GO: GO signal; SR: switch release; OT: object touch; HS: start of hold period; RW: reward). Images above the time axis indicate the state of the 5 LEDs during a grip-first condition at WS, during the presentation of the cue (CUE-ON through CUE-OFF) and at GO. \textbf{B.} Spatial locations of the Utah multi-electrode arrays (green squares) on the cortical surface in monkey L (left), T (middle) and N (right). Top and bottom graphs show the array locations with respect to anatomical features (red curves) estimated from the corresponding photographs shown in panel C. CS: central sulcus; AS: arcuate sulcus; PS: precentral sulcus. \textbf{C.} Photographic image of the array locations taken during surgery. \textbf{D.} Power spectrum of the LFP during the complete recording of one selected central electrode (id 50), averaged across all sessions ($N$=15 per panel) in the grip-first condition of monkey L (left), T (middle), and N (right). Orange shading: range of the applied beta band filter (cut-off frequencies). \textbf{E.} Trial-averaged, time-resolved power spectrogram of the LFPs of one electrode in one recording session during PG trials of a grip-first recording. Trials aligned to TS. Color indicates logarithmic power density. Horizontal dashed lines: beta band as shown in panel D. Vertical dotted lines: trial events (SR, RW: mean times). Session IDs: l101013-002 (monkey L), t010910-001 (monkey T), and i140613-001 (monkey N) from left to right, respectively.}
    \label{fig:spectral-properties}
  \end{minipage}
\end{figure}

\subsection*{Spectral LFP properties}
\label{sec:Results-Spectral-Properties}

On a first glance, we observed that the LFP in all monkeys was dominated by a prominent oscillatory component in the beta range (about \myunits{15 - 35}{Hz}). By computing the average power spectrum of each monkey's LFP, pooled for one electrode in the array center across its complete set of recordings in the grip-first condition (15 per monkey), we found that the frequency range of the beta oscillation varied between monkeys (see \figref{spectral-properties}D). Based on these spectra we defined a wide frequency band (\myunits{13-30}{Hz}) that was common to all monkeys and covered the peaks of the individual beta frequencies (shaded area in \figref{spectral-properties}D). For better comparison, we applied this same filter band in the beta range to all data sets of all monkeys.

Furthermore, the observed LFP activity revealed that the trial-averaged power of the oscillatory activity was not stationary in time, but was strongly modulated during the time course of behavioral trials. The strength of the beta oscillations is visualized by the time-resolved power spectra, averaged on one electrode (same as for \figref{spectral-properties}D) across all successful PG trials of one representative recording session of monkeys L, T and N, respectively (\figref{spectral-properties}E). The beta power showed a characteristic temporal evolution that followed a similar trend for all three monkeys: the beta power was largest around the cue, and decayed gradually during the delay period and was strongly attenuated during movement execution. During movement, a low frequency signal was the most prominent component in the LFP, corresponding to the movement-related potential \citep{Riehle13_48}.

%
%
\begin{figure} \centering
\includegraphics[width=\linewidth]{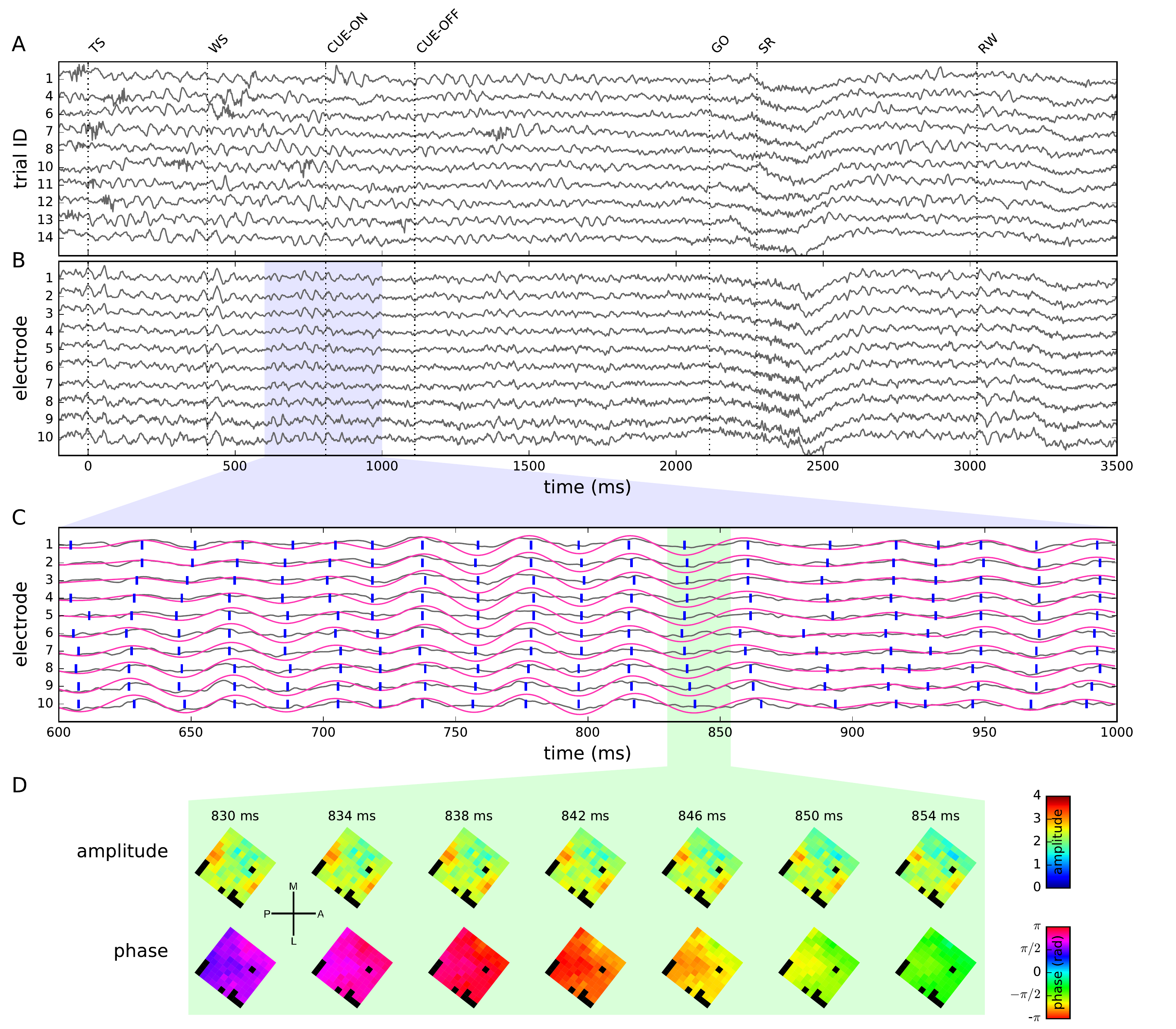} \caption{\textbf{Extraction of phase and amplitude maps}. \textbf{A.} LFPs (z-scored) recorded from one electrode during 10 consecutive successful trials (monkey L, session ID: l101015-001). Trials aligned to TS=\myunits{0}{ms}. \textbf{B.} Simultaneously recorded LFPs from 10 neighboring electrodes on the Utah array during a single trial. \textbf{C.} Blow-up of the LFPs of the 10 example electrodes shown in panel B (gray traces; blue shading in panel B indicates the selected time window). Red traces: beta-filtered LFP. Blue lines: locations of peaks and troughs in the filtered LFP, i.e., phases $\phi=0$ and $\phi=\pi$. \textbf{D.} Amplitude (top) and phase (bottom) maps (shown in \myunits{4}{ms} steps) recorded during a \myunits{24}{ms} window (green shading in panel C). Color in each square indicates the amplitude and phase of the LFP at the electrode of a given position. Black squares: unconnected electrodes or electrodes rejected due to signal quality. The images are rotated to match the cortical position of the array as indicated in \figref{spectral-properties}B.} \label{fig:obtaining-phase-maps}
\end{figure}

The inspection of single-trial LFP signals revealed, in addition to the beta power modulations observed in trial averages, a modulation of the instantaneous amplitude of beta activity (cf.~\figref{obtaining-phase-maps}A) on a much shorter time scale. Such epochs of increased beta activity comprising a few oscillation cycles are commonly referred to as beta spindles \citep{Murthy92_5670,Murthy96_3949}. During a single trial, LFP signals recorded in parallel from all electrodes of the Utah array exhibited in general a high degree of correlation (\figref{obtaining-phase-maps}B), and in particular spindles occurred simultaneously on all electrodes \citep[cf.~also][]{Murthy96_3949}. However, across trials spindles did not reoccur at the same points in time (\figref{obtaining-phase-maps}A), but instead their occurrence in time exhibited a strong degree of variability. Therefore, the trial-averaged temporal evolution of beta power (\figref{spectral-properties}E)  represents a measure that confounds the probability of single-trial high amplitude events, their average duration, and their average maximal amplitude \citep{Feingold15_13687}.

\subsection*{Identification of phase patterns}
\label{sec:Results-Identification-of-Patterns}

Having described the principle properties of the dominant beta oscillations, we are now in a position to investigate the fine spatial patterning of this activity across all electrodes of the array. Zooming in time into the LFP signals recorded from a few neighboring electrodes during the entire trial length (\figref{obtaining-phase-maps}B), we calculated the beta-filtered signals (\figref{obtaining-phase-maps}C, red traces). We observed that despite a high degree of similarity, the oscillatory components express small time lags across the electrodes (compare blue markers on each trace indicating oscillation peaks and troughs). To understand if there is a specific spatial patterning of the temporal lags between the signals on the different electrodes, we decomposed the beta-filtered LFP time series of each electrode $i$ into the instantaneous amplitude $a_i(t)$, corresponding to the envelope of the filtered signal, and phase $\phi_i(t)$ of the beta oscillation by calculating its analytic signal (see \nameref{sec:Methods}). We then displayed these quantities as spatial maps $\ram{xy}$ and $\rpm{xy}$ for amplitude and phase, respectively, representing each electrode at its spatial array position $(x,y)$ at each time point $t$. Even though the oscillation amplitude $\ram{xy}$ was not the same across the array, its modulation was highly correlated between electrodes and showed a pattern that was changing slowly as compared to the time scale of the beta period (\figref{obtaining-phase-maps}D and movie S1 in the Supplemental Information). This finding matches our observation that the occurrence of spindles is coherent across recording electrodes (\figref{obtaining-phase-maps}B).

%
%
\begin{figure}
 \centering
 \includegraphics[width=\textwidth]{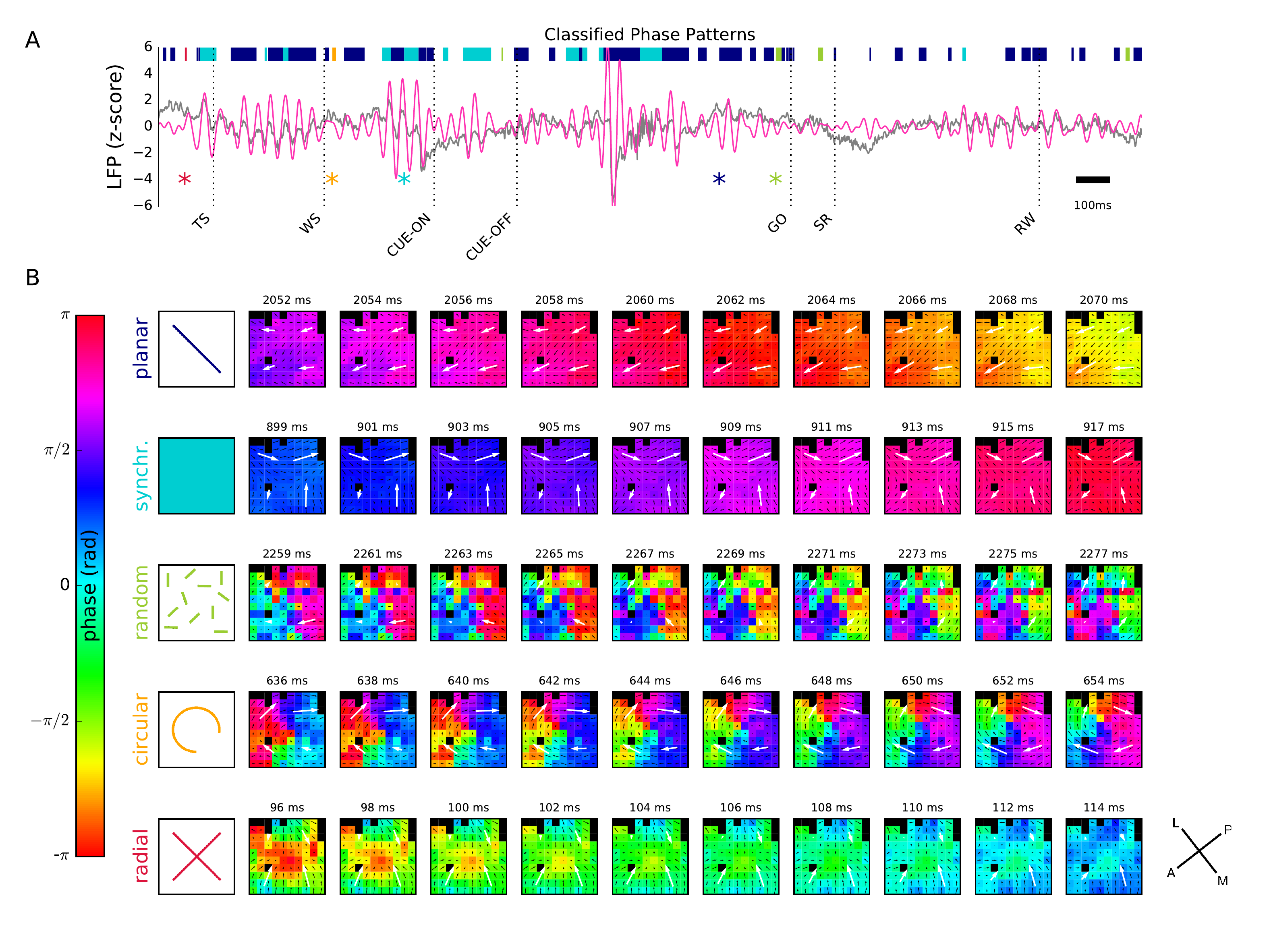}
\caption{\textbf{Phase patterns and their detection}. \textbf{A.} LFP signal (z-scored) recorded during a single trial (monkey L, Session ID: l101108-001, trial ID: 46) on a central electrode (gray) and superimposed beta-filtered LFP (red). Dashed vertical lines indicate trial events. Colored horizontal bars show time periods during which a particular type of phase pattern (compare color code in panel B, first column) was detected. The colored asterisks mark the time points of the first frame of the wave patterns shown in panel B. \textbf{B.} Phase maps of one example of an automatically detected phase pattern for each type of phase pattern (rows, from top to bottom: planar wave, synchronized, random, circular, and radial). The sequence of maps in one row shows a total of \myunits{18}{ms} in steps of \myunits{2}{ms}. The pattern was initially detected in the first phase map of each row (corresponding time point indicated by an asterisk in panel A). Flow field indicated by black lines: gradient coherence map $\gcm{xy}$; white large arrows: corresponding quadrant-averaged gradient coherence shown for visualization. Time stamps are given relative to TS.}
\label{fig:phase-patterns}
\end{figure}

In contrast, the phase snapshots $\rpm{xy}$ showed a pronounced structure that varied on a very fast time scale in the range of milliseconds (\figref{obtaining-phase-maps}D and movie S1 in the Supplemental Information). While we typically observed a smooth transition of maps between consecutive time points $t_i$ and $t_{i+1}$ (given a sampling rate of \myunits{1}{kHz}), at some moments in time the maps changed their structure very rapidly. However, despite the rapid changes of the spatial structure of the maps and some discontinuities in their temporal evolution, many phase maps could clearly be classified by visual inspection into one of 5 distinct classes of spatial arrangements, in the following referred to as \textit{phase patterns}. Representative examples of these classes of phase patterns and their temporal evolution over a time period of \myunits{20}{ms} are shown in \figref{phase-patterns}A,B. In order to better visualize and characterize these spatial structures, we here calculated the vector field of phase gradients $\pgm{xy}$ and its spatially smoothed version, the phase gradient coherence $\gcm{xy}$, and display the gradient fields along with the phase maps. In the following we will briefly describe the classes of phase patterns in their most salient, idealized form.

The identification of traveling waves (\figref{phase-patterns}B, top row), comparable to the first report by \citealt{Rubino06_1549}, was most prominent. In these \textit{planar} patterns, a planar wave front traveled across the array, where the spatial period was typically larger than the array dimensions. Second, we observed a \textit{synchronized} pattern (\figref{phase-patterns}B, 2nd row), in which the signals on all electrodes were synchronized at near-zero phase lag. Complementing this state at the other extreme, we observed a \textit{random} pattern (\figref{phase-patterns}B, 3rd row), which showed no apparent phase relation between electrodes. A fourth pattern, termed \textit{circular} pattern (\figref{phase-patterns}B, 4th row), was characterized by an area near the array center around which the phases revolved. Finally, we observed a \textit{radial} pattern (\figref{phase-patterns}B, bottom row) of radially inward or outward propagating waves, which was also characterized by a point of origin near the array center. A specific type of pattern persisted for only short time periods of approximately  the duration of a single beta oscillation cycle. In addition, some phase maps could not be clearly attributed to one of these 5 phase patterns.

Following this first  empirical identification of classes of phase patterns, we aimed at automatically classifying the sequences of phase maps into one of these classes whenever possible. To this end, we introduced a set of 6 measures that capture features of the spatial arrangement of beta oscillations based on the phase map $\rpm{xy}$, and its spatial arrangement quantified by the phase gradients $\pgm{xy}$ and the gradient coherence $\gcm{xy}$ independently at each time point $t$. The details of how to construct these measures are given in the \nameref{sec:Methods}. Essentially, each of the measures represents a feature of a given phase pattern that is characteristic for one or several of the 5 classes of phase patterns. In the following, we give an intuitive explanation of the features relevant for each individual pattern class. The planar patterns, described by a planar wave front traveling across the entire array, were characterized by phase gradients that were aligned in parallel across the entire array. Thus, such a pattern was composed of a wave front oriented perpendicularly to the gradients. The synchronized pattern was distinguished by a single phase value at all electrodes (i.e., the array appears in a single color in \figref{phase-patterns}B) and a random direction of phase gradients across the array. Similarly, the random pattern showed no apparent local spatial organization of phase gradients, but in contrast phases were uniformly distributed. In the circular pattern, like in the synchronized pattern, phase gradients in all directions were observed, but in contrast the distribution of phases across the array was also uniform such that the visualizations in \figref{phase-patterns}B contained all colors. Additionally, phase gradients were always arranged such that they pointed clockwise or counter-clockwise around the center of the array. And finally, the radial pattern exhibited phase gradients that, on a global view, pointed inward or outward from the array center. Thus, gradients pointed in a direction orthogonal to that of circular patterns. Common to both circular and radial patterns, all phase gradient directions were observed on the array and neighboring gradients on the array were similar.

Based on the 6 measures, we used a thresholding procedure (compare red dashed lines in \figref{phase-patterns-thresholds} in the Supplemental Information) to assign each phase pattern at time point $t$ to one of the 5 classes of patterns, or, if none of the combined threshold criteria was met, the phase pattern was left unclassified. Thresholds were set empirically in such a way that they led to a conservative association of phase patterns with pattern classes, i.e., only clearly identified patterns were classified (see \figref{threshold-setting} in the Supplemental Information for a visualization of accepted and rejected classifications). Details of the classification process are provided in the \nameref{sec:Methods}. Our classification procedure had some experimental limitations such as the low spatial sampling of the 100 electrodes (\myunits{400}{\mu m} inter-electrode distance) and the small spatial window of observation (\myunits{4x4}{mm}) as compared to the spatial wavelengths exhibited by some patterns. This may affect, in particular, the radial and circular patterns in which the point of origin was not at the array center, making it impossible to infer the pattern unequivocally. Additionally, observed patterns could also have represented transient dynamics from switching between patterns or even overlaps of competing patterns, which could not be properly distinguished. If for any of these reasons a pattern did not fulfill the strict criteria of one of the five pattern classes described above, we referred to it as \textit{unclassified}.

The use of our algorithm enabled us to quantitatively disambiguate the 5 phase patterns that appeared as salient features upon visual inspection of the phase maps. The phase patterns shown in \figref{phase-patterns}B were determined by using this algorithm. \figref{phase-patterns}A shows the LFP recorded on one electrode during one single trial, in which all classified phase patterns are marked, including those shown in panel B. The corresponding measures and thresholds used in the classification procedure are depicted in \figref{phase-patterns-thresholds}.

Periods of clearly identified phase patterns were typically of short duration and occurred interspersed throughout the trial. During the entire length of all selected sessions, including both the behavioral trials and the inter-trial intervals, we counted for each monkey the number of occurrences of continuous periods of time where one of the 5 phase patterns or an unclassified pattern was observed. The percentage of time points of the sampled LFP identified as each of the pattern classes is provided in \tabref{Percentage-Classified}. In addition, as a more conservative measure that takes into account potentially spurious patterns that were detected for very brief instances only, the number of epochs of contiguous time points classified as the same pattern and lasting for at least \myunits{5}{ms} is displayed for the grip-first condition in \figref{behavioral-correlates}A (for the force-first condition, see \figref{behavioral-correlates-force-first}A in the Supplemental Information). These results show that all pattern types were observed in each monkey, with planar wave patterns being among the most prominent and circular patterns among the least observed patterns. Only in monkey N, the random pattern was observed more often than the planar wave pattern. In addition, monkey N rarely exhibited a synchronized pattern as compared to monkeys L and T.

\begin{center}
\begin{table}
\caption{\textbf{Percentage of time points classified as a specific phase pattern} in each monkey given the conservative choice of thresholds used in the analysis (pooled over all grip-first and force-first conditions).}
\ \\
\label{tab:Percentage-Classified}
\begin{tabular}{|l|l|l|l|}\hline
\ & \multicolumn{3}{|l|}{\textbf{Monkey}} \\\hline\hline
\textbf{phase pattern} & L & T & N\\\hline\hline
planar       & 28.47 & 42.41 & 5.51\\\hline
synchronized & 1.81  & 14.98 & 0.01\\\hline
random       & 1.28  & 0.05  & 8.93\\\hline
circular     & 0.01  & 0.01  & 0.02\\\hline
radial       & 1.14  & 0.46  & 1.47 \\\hline
unclassified & 67.28 & 41.65 & 84.06 \\\hline
\end{tabular}
\end{table}
\end{center}

\subsection*{Relation of beta amplitude and phase patterns to behavioral epochs}
\label{sec:Relation-Behavior}

Given the abundance of patterns in the data, we asked whether there is a relationship between phase patterns and behavioral epochs of a trial. Thus, we investigated whether or not the occurrence of a specific phase pattern is linked to one or more behavioral events. We determined trial by trial and for each pattern separately its precise occurrence during the time course of the behavioral trial. We pooled the data from each monkey across all trials of the same condition (correct trials only), i.e.\ grip-first or force-first condition, and across all selected recording sessions, to obtain a measure for the probability of the occurrence of a specific pattern in time. In the following, we discuss in detail data from the grip-first condition, but qualitatively similar results are seen in the force-first condition (see \figref{behavioral-correlates-force-first} in the Supplemental Information). \figref{behavioral-correlates}B shows that monkey N had comparatively low numbers of planar and synchronized patterns during the trial, but a higher number of random patterns than the two other monkeys. This suggests that many of the planar and synchronized patterns of monkey N observed during the complete recording (\figref{behavioral-correlates}A) occurred during the inter-trial intervals, and not during the trial (\figref{behavioral-correlates}B).

%
%
\begin{figure}
 \centering
 \includegraphics[width=\textwidth]{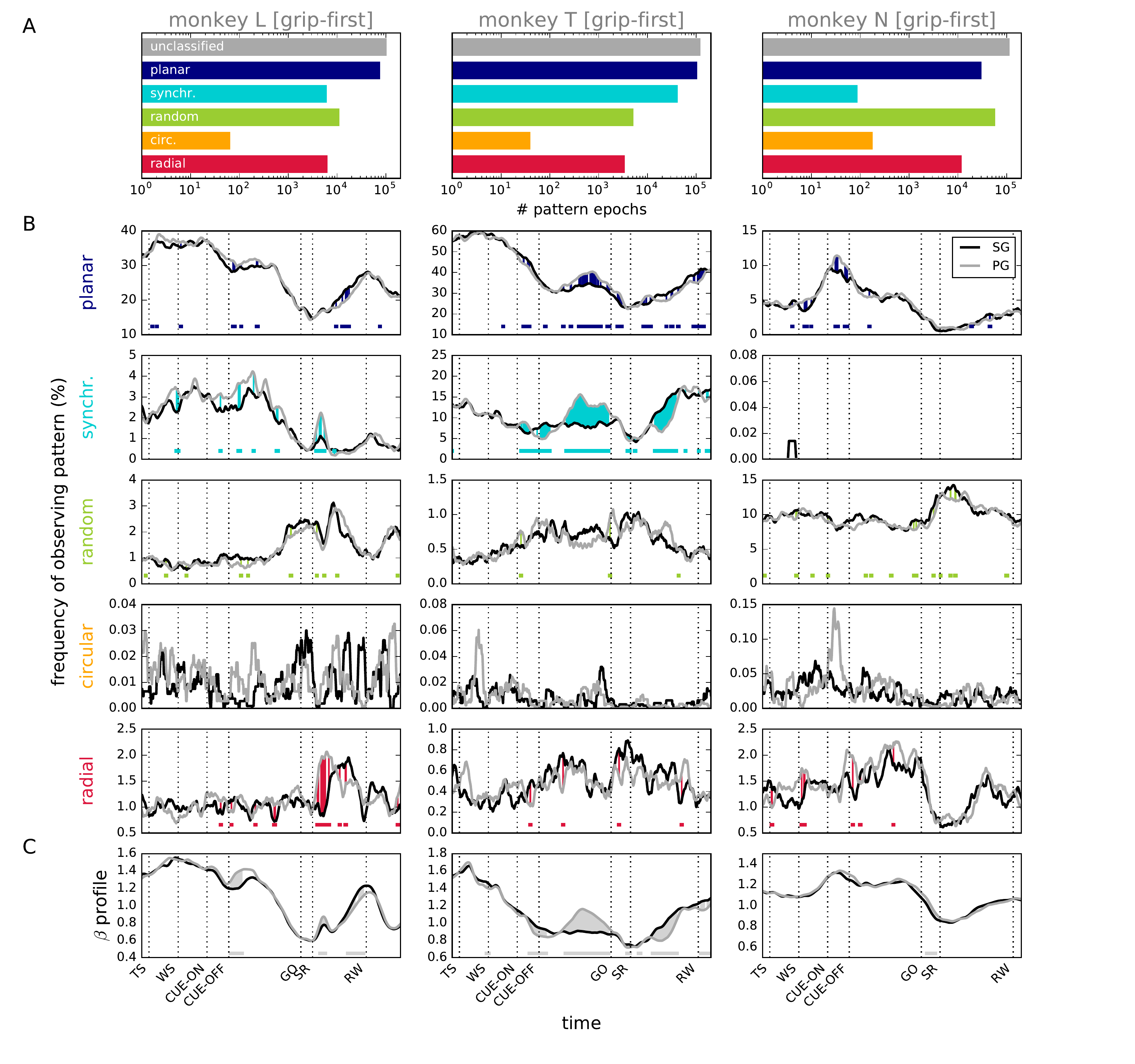}
 \caption{\textbf{Behavioral correlates and relation to average beta amplitude for the grip-first condition.} \textbf{A.} Number of epochs of a phase pattern detected in at least 5 consecutive time frames, i.e., \myunits{5}{ms} (bars from top to bottom: unclassified, planar wave, synchronized, random, circular, radial pattern) for monkey L (left), T (middle), and N (right). Data were obtained from all selected recording sessions including inter-trial intervals. \textbf{B.} Time-resolved probability of observing a specific phase pattern (rows) during the trial. Statistics were computed across all grip-first trials of all recording sessions for each monkey ($N=15$) and smoothed with a box-car kernel of length $l=$\myunits{100}{ms}. Trials were separated into side-grip (SG) trials (black) and precision-grip (PG) trials (gray). For monkey N, only very few synchronized patterns were detected during the trial. Color shading between curves and colored bars indicate time periods where SG and PG curves different significantly (Fisher's exact test under the null hypothesis that, for any time point, the likelihood to observe a given phase pattern is independent of the trial type, $p<0.05$). \textbf{C.} Beta amplitude profile (envelope) pooled across all SG (black) and PG (gray) trials (same data as in panel B). The amplitude profile $\amplitude$ of a single trial is calculated as the time-resolved instantaneous amplitude $\ram{xy}$ of the beta-filtered LFP averaged across all electrodes $(x,y)$, and measures the instantaneous power of the beta oscillation in that trial. Gray shading between curves and horizontal bars indicate time periods where SG and PG curves differ significantly (t-test under the null hypothesis that the distributions of electrode-averaged single trial amplitudes $\ram{xy}$ at each time point $t$ are identical for SG and PG trials, respectively, $p<0.05$).}
 \label{fig:behavioral-correlates}
\end{figure}

In the next step, we assessed similarities in the temporal profile of the pattern occurrence probabilities during the behavioral trial (\figref{behavioral-correlates}B). For each monkey, the probability of observing any pattern was strongly modulated over the time course of the trial. Common to all monkeys was the finding that planar patterns occurred mostly during the initial cue presentation and during reward, and were less prominent during movement. Synchronized patterns expressed a similar time course for monkey L and to a lesser degree for monkey T. Monkey N showed almost no synchronized pattern during the trial. In contrast, in all monkeys random patterns occurred predominantly towards the end of the delay period and during movement. Circular and radial patterns were rarely observed during the trial , but exhibited a clear modulation structure in time , albeit in a different way for each monkey.

The specific and consistent temporal modulation of the occurrence probability suggests that the spatio-temporal structure of activity is related to motor cortical processing performed during the trial. We thus asked, if also particularities of the trial condition were reflected in the probability. To this end, we compared results obtained during SG and PG conditions (\figref{behavioral-correlates}B, black and gray, respectively). In general, the modulations of probability for both trial types were very similar, but expressed a few notable exceptions. For planar waves, SG and PG deviated slightly, but significantly (indicated by dots at the bottom of each panel), during early delay (probability of observing a pattern during PG trials exceeded that of SG trials, PG$>$SG) and before reward (PG$<$SG) for monkey L, during late delay (PG$>$SG) in monkey T, and during cue presentation for monkey N (PG$>$SG). Similar, even more pronounced observations were made for synchronized patterns of monkey L and T. In addition, a tendency for a reversed effect was observed for random patterns in particular during the delay period of monkeys L and T.

Up to now, we concentrated on the time-resolved spatial organization of oscillatory activity on the basis of the phase information extracted from the time series. We next asked how these findings relate to the trial-averaged beta power, because we noticed that the temporal evolution of the occurrence probability of planar and synchronized phase patterns was reminiscent of the evolution of the beta power (\figref{spectral-properties}E).
To further investigate this observation, we calculated the trial-averaged beta amplitude profiles $\amplitude$, i.e., the time-resolved instantaneous amplitude, or envelope, $\ram{xy}$, of the beta signal pooled across all electrodes $(x,y)$, as a representative of the average instantaneous power of the beta oscillation. Again, data were calculated for all sessions used in \figref{behavioral-correlates}B and separately for SG and PG trials (\figref{behavioral-correlates}C). Interestingly, for all 3 monkeys the time-resolved beta amplitude profiles closely followed the occurrence probability of the planar phase pattern (\figref{behavioral-correlates}B, top). For monkeys L and T, also the time course of synchronized patterns loosely followed that of the beta amplitude profiles. In particular, we noticed that all differences between SG and PG trials identified in the pattern occurrence probabilities were reflected in the beta amplitude profile. For example, in monkey L, the beta profiles obtained in SG and PG trials differed during the early delay (PG$>$SG) and before reward (PG$<$SG), mirrored in the occurrence of planar waves, and after movement onset (PG$>$SG), mirrored in the occurrence probability of synchronized patterns. Similar observations were made for the other monkeys, and in the time period after the GO signal for the force-first condition (\figref{behavioral-correlates-force-first}).

\subsection*{Quantification of phase patterns}
\label{sec:Quantification-of-patterns}

So far our findings revealed that the modulation of the probability to observe any of the 5 phase patterns, both in time and by the behavioral condition, was correlated with the average beta amplitude. This suggests that the spatial organization of oscillatory activity, represented by the different phase patterns, may not only be reflected in the trial-averaged power in a statistical sense, but that indeed amplitude modulations of the oscillatory LFP, and in consequence also beta spindles, correlate with the patterns on a single trial level.

As a first step to understand the properties of phase patterns in single trials, we quantified features extracted from the classification results. As classification was performed on single time points, we first calculated the durations of epochs of consecutive time points being classified as the same pattern. In \figref{statistical-properties}A we show the resulting distributions of the durations for each of the pattern types. Naturally, these statistics depended on how conservative the choice of thresholds for pattern detection was set. However, given that thresholds were set in accordance to the observed phase pattern (cf.~\figref{threshold-setting}), they served as a visually inspired characterization of the observed duration of a pattern. We found that, on average, planar, synchronized, and radial patterns all had longer durations than random and circular patterns (see large dots in \figref{statistical-properties}A), all on the order of less than one cycle of the beta oscillation (\myunits{\approx 40 - 50}{ms}).

%
%
\begin{figure}
 \centering
 \includegraphics[width=\textwidth]{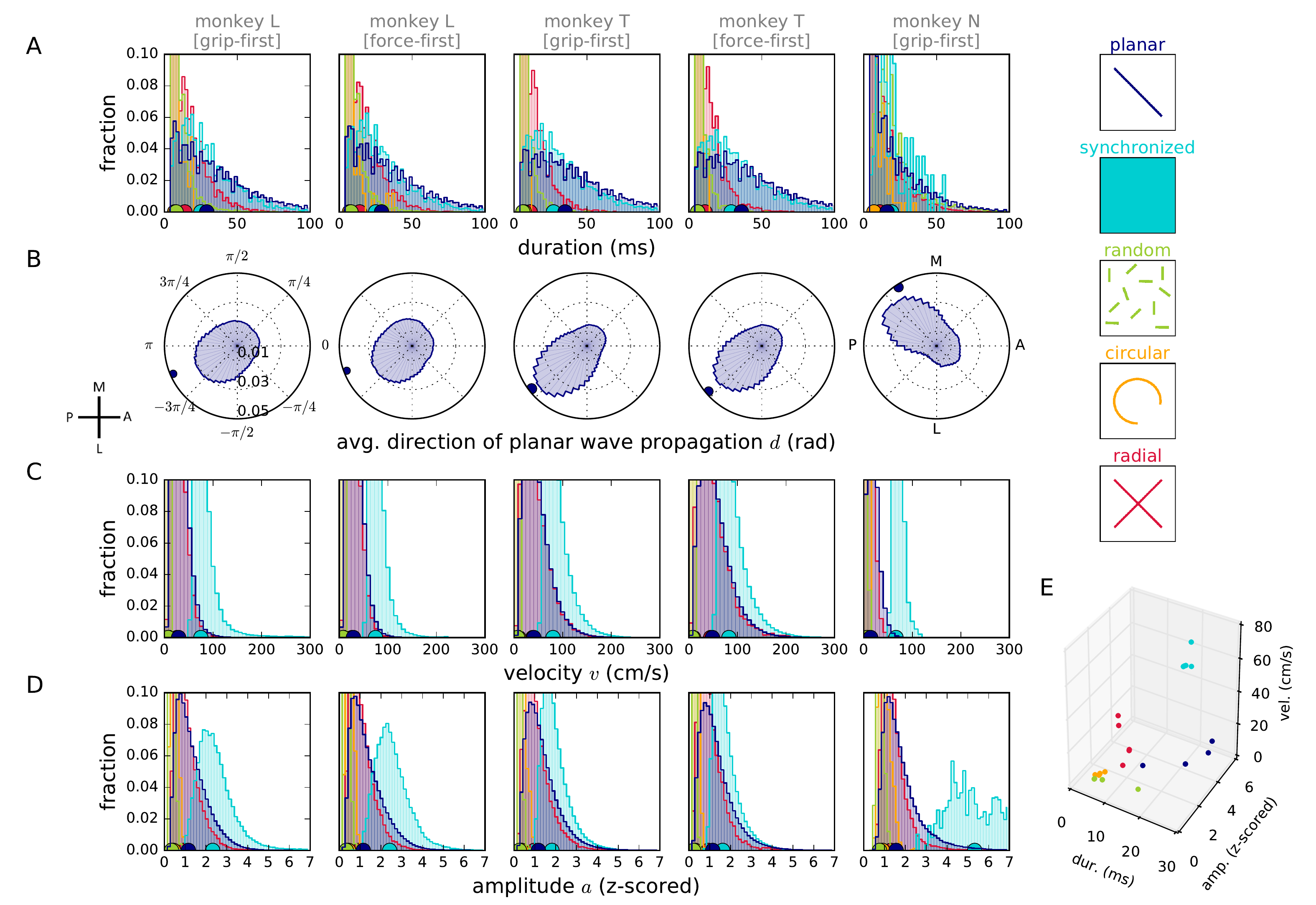}
 \caption{ \textbf{Statistics of detected patterns.} \textbf{A.} Histogram of durations of epochs of consecutive time points classified as belonging to the same phase pattern (cf.,~\figref{behavioral-correlates}A). \textbf{B-D.} Distributions of the direction $\direction$ (panel B), phase velocity $\velocity$ (panel C), and amplitude profile $\amplitude$ (panel D), as a function of the detected phase pattern. Data are separated (columns) according to monkey and recording  condition (grip-first vs.\ force-first). Histograms for different phase patterns are plotted overlapping in the color corresponding to the legend on the right. For each phase pattern, a histogram entry in panels B-D represents the measured quantity averaged across the array calculated at a time point classified as that pattern. In panel B, the average direction of the phase gradient is plotted in brain coordinates by rotating the activity, and mirroring data along the medio-lateral axis for monkey T to compensate for the array placement in the opposite hemisphere as compared to L and N. Large semi-circles: medians of the corresponding distributions. \textbf{E.} Joint representation of the medians of the distributions shown in panels A, C, and D. Each data point represents the median of one monkey in one recording condition for one pattern class (indicated by color).}
 \label{fig:statistical-properties}
\end{figure}

In a next step we examined the preferred direction of the phase gradients of the wave patterns. Here we only considered planar phase patterns (\figref{statistical-properties}B) for which the measure was equivalent to the direction of movement of the planar wave front. Planar waves in monkeys L and T were preferentially observed in the anterior-medial to posterior-lateral direction (see inset for cortical space), whereas waves in monkey N were observed in the anterior-lateral to posterior-medial direction. Noting that the array location in monkey N differed from that in monkeys L and T (cf.~\figref{spectral-properties}B,C), the observations from all three monkeys are compatible with those described in \citealt{Rubino06_1549}. Even though it was possible to calculate the direction of the phase gradients of any phase pattern, we refrained from showing the distribution of directions for patterns other than planar patterns since their characteristics do not allow a clear interpretation of wave propagation direction.

To investigate the dynamical aspect of wave propagation, we calculated the average wave velocity $\velocity$  (cf.~\nameref{sec:Methods}) at each time point (\figref{statistical-properties}C). For planar wave patterns, this was directly interpretable as the propagation velocity of the observed wave front. The median propagation velocities of the planar waves were $\velocity=$\myunits{29.1\pm 10.3}{cm/s} (grip-first) and $\velocity=$\myunits{29.1\pm 10.4}{cm/s} (force-first) for monkey L, $\velocity=$\myunits{40.5\pm 16.2}{cm/s} (grip-first) and $\velocity=$\myunits{40.3\pm 19.5}{cm/s} (force-first) for monkey T, and $\velocity=$\myunits{14.2\pm 4.6}{cm/s} (grip-first) for monkey N (all values: median $\pm$ median absolute deviation). These values are in rough agreement with those reported in \citealt{Rubino06_1549}. For the other wave patterns, even though it was possible to calculate a velocity, it may not be directly interpreted as the velocity of a propagating planar wave front since phase gradients do not align across the array. Instead, it is a measure that captures the average velocity calculated from the local velocities across the array. Synchronized patterns could be considered as a special case of planar waves with a very large spatial wavelength, and as a consequence they exhibited high (in theory, infinitely high) velocities (\figref{statistical-properties}C). On the other hand, random and circular patterns were characterized by phase values that differed strongly between adjacent recording sites. Therefore, average velocities derived from the phase gradients were low. Finally, radial patterns resembled the planar patterns in that they could be approximated by a planar wave front at a large distance from the center of the radial pattern. In agreement with this interpretation, they exhibited similar phase velocities as observed for the planar pattern. Thus, the 5 distinct phase patterns showed clear differences in the distribution of their velocities, where a low $\velocity$ corresponds to random or circular patterns, a medium $\velocity$ relates to planar or radial patterns, and a high $\velocity$ indicates the presence of a synchronized pattern. In this sense, the value of the velocity represents a reliable proxy for the type of the observed pattern.

After having quantified the features of the wave patterns in single trials, we come back to the question of how the beta amplitude relates to the occurrence of a wave pattern. In \figref{statistical-properties}D we show for each monkey and behavioral condition the distribution of the beta amplitude profile (envelope) $\amplitude$ during each of the 5 phase patterns. As predicted above from trial-averaged data, we observed a clear relationship between the instantaneous magnitude of the beta oscillation and the spatial phase pattern. Circular and random patterns occurred at small amplitudes, planar and radial patterns at intermediate amplitudes, and only synchronized patterns occurred at high amplitudes. Therefore, the approximate correspondence between the probability of observing a pattern and beta amplitude seen in the trial average (\figref{behavioral-correlates}) is consistent with the relationship between the single-trial amplitude modulation of the beta oscillation and its spatial organization given by the phase pattern.

The results shown in \figref{statistical-properties}A, C and D were summarized in \figref{statistical-properties}E, where for each phase pattern, monkey and behavioral condition the averaged data for duration, velocity and amplitude are plotted against each other. This representation clearly shows a clustering of collective data points for each individual phase pattern. Thus, the 5 phase pattern classes are  described by a specific combination of characteristic values for pattern duration, velocity, and amplitude.

\subsection*{Beta amplitude determines phase pattern}
\label{sec:Amplitde-Determines-Pattern}

%
%
\begin{figure}
 \centering
 \includegraphics[width=\textwidth]{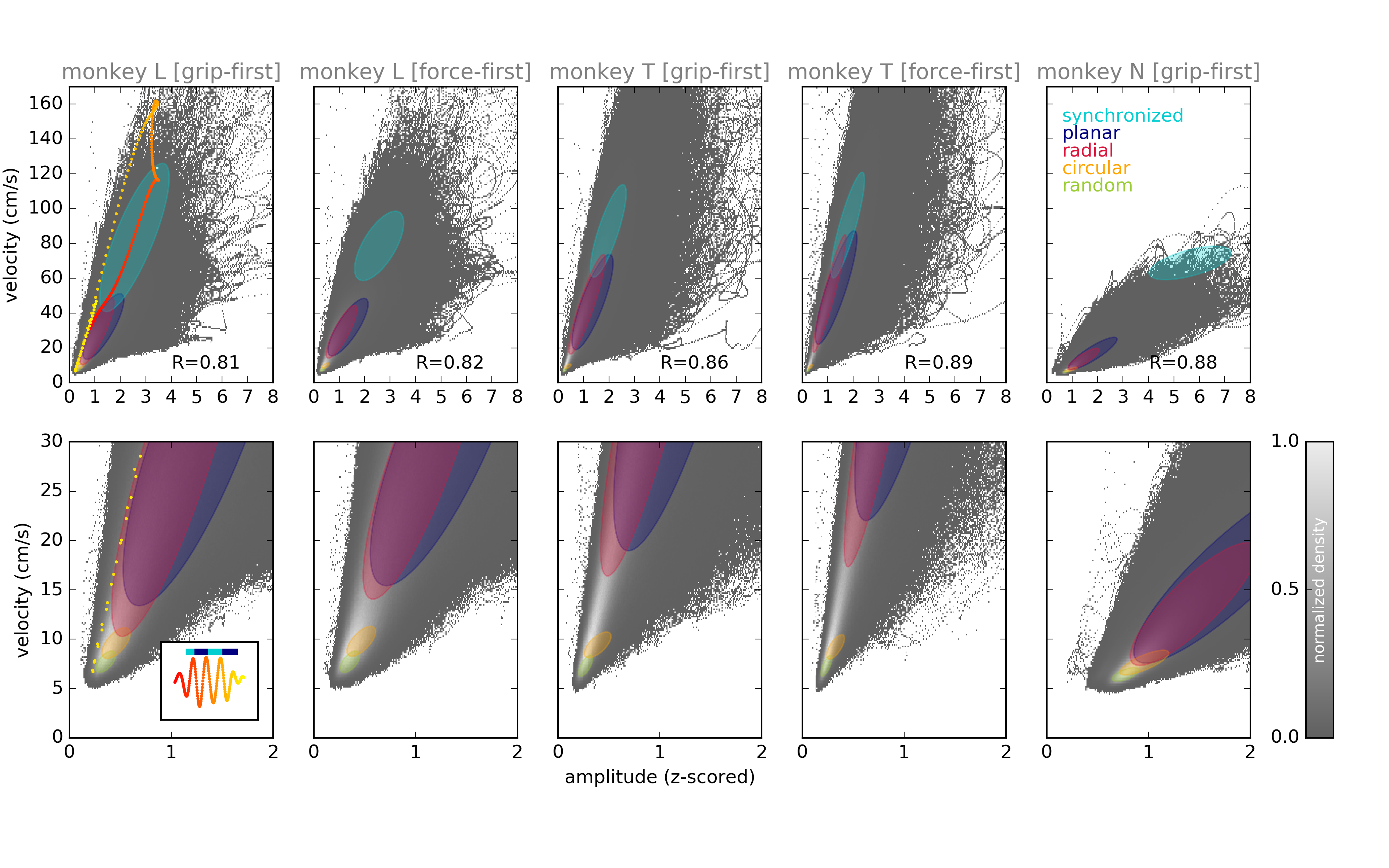}
 \caption{\textbf{Correlation of instantaneous beta power and spatial pattern of phases.} Upper row: 2-D histograms of phase velocity $\velocity$ and beta amplitude profile $\amplitude$ evaluated at each time point (independent of the detected phase pattern) shown for each monkey and behavioral condition (columns). Gray values indicate density of time points falling in each histogram bin, normalized to the largest entry of the histogram. The values of the Pearson correlation coefficients $R$ are given in the bottom right of each panel. Each ellipse represents the distribution of time points classified as a specific phase pattern (indicated by color). Center of ellipses: mean; radii of ellipses are given as 2 standard deviations in the direction of the 2 principle components. Lower row: zoomed-in versions of the upper histograms. Left column: Inset in lower panel and dots in red to yellow shades: Illustration of spindle dynamics by example of the spindle before CUE-ON presented in \figref{phase-patterns}A. Inset reproduces this spindle (transition from red to yellow colors indicates increasing time), corresponding detected states are shown as bars above the spindle. For each time point of the spindle, the corresponding values of the amplitude and phase velocity are marked in the histograms using the identical color. Average ellipse centers: \myunits{(1.4\pm0.8}{}, \myunits{41.5\pm27.3}{cm/s)} for planar; \myunits{(1.9\pm0.8}{}, \myunits{88.1\pm29.4}{cm/s)} for synchronized; \myunits{(0.7\pm0.2}{}, \myunits{6.7\pm1.0}{cm/s)} for random; \myunits{(0.7\pm0.3}{}, \myunits{8.8\pm1.7}{cm/s)} for circular; \myunits{(1.2\pm0.7}{}, \myunits{26.8\pm21.5}{cm/s)} for radial.}
 \label{fig:amplitude-velocity-corr}
\end{figure}

In the last step of our analysis, we now ask if the relationship between amplitude and spatial organization holds for any time point independent of whether or not it can be unambiguously attributed to any of the idealized classes of phase patterns. In order to obtain such a time-resolved view of how the amplitude (which by itself did not exhibit a strong spatial organization, cf.~\figref{obtaining-phase-maps}D) correlates with the temporal evolution of the patterns, we employed the phase velocity $\velocity$ as a proxy to quantify the spatial organization that can be readily calculated for each individual time point (as opposed to pattern duration). In \figref{amplitude-velocity-corr} we show the correlation between the instantaneous beta amplitude profile (envelope) $\amplitude$ and the phase velocity $\velocity$ for each time point for all three monkeys, independent of the phase pattern classification, thus including instances during which no pattern could be classified by our conservative classification algorithm. We observed that the two variables were very strongly positively correlated ($R>0.8$ for all monkeys) and correlations were highly significant ($p\ll 0.001$). Thus, an increase in amplitude goes along with an increase in phase velocity (cf.~also \figref{statistical-properties}C). As shown above, the velocity $\velocity$ is indeed a good correlate of the perceived organization of beta activity on the electrode array. To more directly illustrate how the velocity measure relates to the previously defined classes of phase patterns, we indicate in \figref{amplitude-velocity-corr} by ellipses the regions of the histograms where the individual classes of phase patterns were predominantly found. In conclusion, we find that at any point in time, the amplitude of the beta oscillation at one single recording site of the Utah array is highly predictive of the spatial organization of activity across the array, here parametrized by the velocity. 

\end{results}

%
%
\begin{discussion}
\label{sec:Discussion}
\pdfbookmark[0]{\discussiontitle}{sec:Discussion-pdf}

Three main objectives guided this work. First, we aimed to obtain a more complete description of the wave-like spatio-temporal phase patterns exhibited in the  beta range of the LFP signals in monkey motor cortex during a complex delayed motor task, and thereby extend reports that only included descriptions of planar wave propagation \citep{Rubino06_1549,Takahashi15_7169}. Second, we aimed at relating the phase patterns to behavioral epochs to determine their possible functional implications. Third, we asked in how far these patterns, determined solely by the phase of the oscillation, are related to the instantaneous modulation of the beta amplitude.

\subsection*{Motor cortical beta oscillations exhibit a variety of spatio-temporal patterns }

By analyzing the dynamics of LFP activity across multi-electrode arrays, we demonstrated that beta oscillatory activity shows a number of salient types of spatio-temporal patterns in addition to traveling planar waves \citep{Rubino06_1549}, namely quasi-synchronized, random, radial, and circular patterns. Such additional types of patterns have previously been predicted from theoretical considerations \citep[e.g.][]{Ermentrout01_33}, and were observed in experimental work, e.g., in slow delta activity of anesthetized marmoset monkeys \citep{Townsend15_4657}. We developed a phenomenological classification method to identify epochs that unambiguously exhibit one of the 5 pattern classes. Our approach detected those in a very conservative manner in order to capture the qualitatively salient patterns that are also identified by a human observer. Indeed, the algorithm tends to leave a large number of time points unclassified, due to the difficulty to clearly attribute a pattern to one of the 5 idealized pattern types. The reason for this is two-fold: On the one side, the coarse-grained resolution of the Utah array  provided only rough estimates of the phase gradients. On the other side, the patterns were often ambiguous, in particular at time points of dynamical transitions between patterns. Planar wave fronts were often not completely planar, but showed a slight curvature, a feature shared with radial or circular patterns. Furthermore, radial and circular patterns that were not necessarily centered on the array were difficult to detect. Also, random states often exhibited a slight degree of correlation between activities recorded on neighboring electrodes, contradicting the \textit{a priori} assumption of pure independence. Nevertheless, this approach of detecting patterns yielded reliable results in terms of their statistics (\figref{statistical-properties}).

To overcome the limitation that the phenomenological classification method only detected unambiguous phase patterns, we tested the potential of the phase velocity as an easily accessible continuous measure to quantify the spatial arrangement of phases for time points where none of the ideal pattern categories matched the observation. Due to the fact that the velocity vector is tightly coupled to the arrangement of phase gradients across the array (see \figref{statistical-properties}C,E), we could indeed link the distributions of velocities to the 5 specific phase patterns (see \figref{statistical-properties}E). Thus, using the continuous measure of the phase velocity, we were able to gain a complete picture of the time course of pattern progression.

The instability of pattern types may suggest that some of the salient pattern types indeed underlie identical dynamical processes, and form a continuum: radial patterns may appear nearly planar wave-like some distance from the array center, and quasi-synchronized states appear at the limit of planar waves approaching infinite phase velocity. This similarity of phase patterns was also reflected in the statistics (\figref{statistical-properties}) describing the occurrence of the patterns, e.g., the similar duration of radial and planar patterns, or the comparable distributions of velocity for planar and radial, as well as circular and random patterns. To investigate this issue in detail, recordings on a larger spatial scale and with a higher spatial resolution would be required.

\subsection*{Specific phase patterns occur at different times during movement preparation and execution}

The probability of detecting a specific phase pattern was variable during the trial of our reach-to-grasp task. Planar and synchronized patterns occurred more often during the pre-cue epoch and during the delay whereas random patterns were more likely to occur around movement execution (\figref{behavioral-correlates}). This observation is in line with the hypothesis that planar and synchronized patterns could be triggered by the arrival of visual information in motor cortex from adjacent cortical areas not covered by our Utah array \citep{Takahashi15_7169}. In agreement with this view, the orientation of planar wave propagation in our data is in agreement with previous studies \citep{Rubino06_1549}. More precisely, we found orientation preferentially aligned along the antero-posterior axis. The direction of planar wave propagation was more anterior-medial to posterior-lateral in monkeys L and T, whereas in monkey N it pointed from anterior-lateral to posterior-medial. This difference could reflect the fact that the array was implanted more medially in monkeys L and T than in monkey N (\figref{spectral-properties}). Therefore, it seems that planar waves travel toward a medial point along the central sulcus, probably at the level of the hand and finger representation ("nested organization", \citealt{Kwan78_1120}; "horseshoe" structure, \citealt{Park01_2784}). This directional preference may be structured by the underlying connectivity of this cortical area \citep{Kwan78_1120}.

The predominance of random patterns during movement execution suggests that the spatio-temporal dynamics of neuronal activity is strongly altered during this epoch. The spatio-temporal structure of these patterns  characterized by their focal origin and short-range propagation could reflect  that during movement, information processing is more local and activity propagation is spatially constrained to motor cortex. However, this hypothesis can hardly be tested at the restricted spatial scale of a single Utah array. Multiple Utah arrays or optical imaging techniques are required to measure the neuronal dynamics at the mesoscopic scale  (see \citealt{Muller14_3675}, for visual cortex).

\subsection*{Wave dynamics relate to the modulation in beta amplitude}

Beta amplitude is known to be strongly modulated by the task epoch \citep{Kilavik13_15}. Interestingly, \figref{behavioral-correlates}  suggests that across trials, the probability of observing different phase patterns also follows the trial-averaged amplitude profile of the beta oscillation. Namely, planar and synchronized waves are present during epochs of large beta amplitudes whereas random waves are prominent during epochs of small amplitudes. The relation  of circular and radial patterns to the beta amplitude is more ambiguous. These observations would  support the hypothesis that the wave dynamics is closely linked to the processes underlying the modulation of the amplitude of beta oscillations.

Indeed, even on the single trial level, \figref{statistical-properties} suggests that low beta amplitudes are linked to random or circular phase patterns with low velocities, intermediate amplitudes to planar or radial phase patterns with intermediate velocities, and that the highest amplitudes indeed co-occurred with quasi-synchronous phase patterns expressing by far the highest velocities (see especially \figref{statistical-properties}E). The pattern statistics also show that the epochs during which one particular, clearly structured pattern was observed were typically of very short duration, in the order of 1 or 2 oscillation cycles (see \figref{statistical-properties}A). This is reminiscent of the short-lasting high amplitude events of a few cycles of beta oscillations described by others, the so-called spindles \citep{Murthy92_5670,Murthy96_3949}. Indeed, these observations point to a tight relationship between spindle dynamics describing the amplitude modulation of the LFP, and the occurrence of wave-like activity, as shown by the correlations  in \figref{amplitude-velocity-corr}. In all monkeys, we observed that with growing amplitude wave propagation tended to accelerate. For high amplitude beta signals, the phase pattern accelerated to such high levels that the observed pattern became synchronous, which in the ideal case would exhibit infinite velocity.

To illustrate how this observation relates to the dynamics of a single spindle, we visualized the temporal evolution of one example spindle and its pattern classification in the left column of \figref{amplitude-velocity-corr} (inset) and observed the corresponding smooth trajectory in the space of amplitude and phase velocity (yellow-red trace). In this spindle, a synchronized state was detected at the spindle peak, flanked by planar patterns before and after the peak. By observing the trajectory of the phase velocity, we observe that the modulation of spindle activity goes along with wave-like activity that progressively increases in speed as the LFP beta amplitude increases. Thus, this strong correlation between amplitude and velocity suggests that at the spindle peak also the velocity peaks, which, for large spindles, corresponds to large spatial wavelengths of the phase pattern that are perceived as synchronized states on the spatial scale of a Utah array. In contrast, a low LFP beta amplitude, as observed between spindles, goes along with random or circular patterns at low velocity. A dynamic representation of how the pattern velocity follows beta amplitude and the evolution of spindles can be seen in the middle panels of the movie S1 in the Supplemental Information.

In summary, we speculate that the formation of a structured, directed pattern, its acceleration to a near-synchronized appearance, followed by deceleration, and finally its breakup in a random or circular pattern marks the temporal organization of the formation of a beta spindle, its peak, and its decay, respectively. Supporting this view, it has been shown that the maxima of LFP spindles tend to synchronize across large distances, even between cortical areas and hemispheres \citep{Murthy92_5670,Murthy96_3949}, as expected for emergence of synchronized patterns. These combined observations are in line with the highly dynamic nature of pattern occurrences.

A model of brain processing that would be intrinsically affected from such a dynamic scaffold is the concept of \textit{communication through coherence}, proposed by \citep{Fries05_474,Fries15_220}. In this framework, the coherence and phase relationship between oscillations on different electrodes were taken as a measure of the ability of neurons to communicate, i.e., that information is best transmitted when the two communicating sites exhibit an optimal phase lag. This concept seems evident when considering, e.g., communication between two brain areas that exhibit distinct population oscillations. It is, however, unclear what this model implies on the mesoscopic scale, such as the course-grained recordings from a Utah array presented here, where the overall pattern of these phase lags between electrodes continuously changes in time. Nevertheless, we may hypothesize that if activities on different electrodes become increasingly synchronized with a decreasing phase lag as spindles increase their amplitude, this would lead to a state where information can be more easily communicated across the complete array, although possibly with less specificity. This would indicate that amplitude modulations, and in particular spindles, act as a time window for enabling  cortical communication across larger distances: not just by means of the strength of synchronization within the local population of neurons (as indicated by the increased beta amplitude), but because this goes along with a wide-spread zero-lag synchronization of the oscillatory activity, i.e., synchronized patterns \citep{Murthy92_5670,Murthy96_3949}. This assumption is highly consistent with the above-described results showing that synchronized and planar patterns are more frequent during the delay epoch and could reflect the transmission of information between distant cortical sites. Conversely, the random patterns occur more often during periods of low beta amplitude and could be linked to the local processing of information. Radial and circular patterns occupy an intermediate position along this continuum and have an unclear relationship to behavior.

This line of arguments raises the question how the patterns of phase dynamics are related to synchronization on the level of single neuron spiking activity. Indeed, it has been shown that the spiking activity synchronizes with the oscillatory spindle peaks \citep{Murthy96_3949} and the cross-correlation histograms of the spiking activity of pairs of neurons become oscillatory in the beta range during periods of strong beta activity \citep{Murthy96_3968}. The entrainment of single neuron spiking activity to the LFP oscillation increases with LFP amplitude \citep{Denker07_2096}. Additionally, we have  shown \citep{Denker11_2681} that at moments of precise transient spike synchronization  that exceeds the expectation based on firing rate \citep{Riehle97_1950}, spikes lock  more strongly to the LFP beta oscillation than expected by chance. This effect of particularly strong locking of significant spike coincidences was observed especially during high beta amplitudes. Interpreting the occurrence of excess synchrony as reflecting active cell assemblies, we embedded our findings in a theoretical model that predicts that activated cell assemblies are entrained to the LFP oscillation at a specific phase shortly preceding the trough of the oscillation \citep{Denker10_599}. Combining these findings, we may speculate that the modulation of the beta amplitude as a function of the occurrence of a beta spindle is not only indicative of the spatial phase pattern of LFP beta activity, as shown in this study, but that additionally beta spindles may govern the temporal structure of spike patterning observed across the array. Indeed, findings of spike sequences \citep{Takahashi15_7169} or synchronous spike patterns \citep{Torre16_8329} that align to the principle direction of phase gradients (\figref{statistical-properties}B) support this view of a functional mechanism that underlies the generation of beta phase patterns.

The discussion by \citet{Muller14_3675} of the functional implications of wave propagation is highly related to such an hypothesis, despite the qualitative differences in their description of waves in superficial layers of visual cortex. The authors show that a network of excitatory and inhibitory neurons operating in the balanced regime and connected by a horizontal fiber network captures the essential features of the observed wave dynamics. Thus, the authors speculate that the transient depolarization caused by the wave passing at a certain position creates a time window of increased sensitivity, i.e., spike probability, of neurons at that location. This would ensure an optimal integration of information as long as the incoming input is timed to the arrival of the wave. Translating this idea to our scenario, the continuous traveling waves we observed could play a similar role when perceived as reverberating waves of the single-cycle propagation \citet{Muller14_3675} have observed. Moreover, we propose that the synchronized patterns, and thus epochs of large beta amplitudes, correspond to states where the optimal time window for the integration of incoming inputs is no longer spatially modulated by the propagating wave dynamics, but only by the anatomical structure of the network.

In summary, despite the fact that motor cortical beta oscillations show a strong correlation between signals recorded over large distances, the phase relationships are highly correlated to the amplitude modulation of beta activity, which in turn has been related to the dynamics of spike synchronization, and to behavior. Thus, we believe that the investigation of amplitude and phase patterns provides a novel leverage on understanding the coordination of activity within spiking neuronal networks.
\end{discussion}

%
%
\begin{methods}
\label{sec:Methods}
\pdfbookmark[0]{\methodstitle}{sec:Methods-pdf}

\subsection*{Experimental Design}
\label{sec:Methods-Experimental-Design}
Three monkeys (Macaca mulatta) were used in these experiments, two females (monkeys L and T) and one male (monkey N). All animal procedures were approved by the ethical committee of the Aix-Marseille University (authorization A1/10/12) and conformed to the European and French government regulations. Monkeys were kept in colonies of 2-4 monkeys in a modular housing pen, with access to a central play area. They were not water-deprived during the experimental period. Each monkey was trained to grasp, pull and hold an object with low force (LF) or high force (HF) using either a side grip (SG) or a precision grip (PG). The task was programmed and controlled using LabView (National Instruments Corporation, Austin, TX, USA). The trial sequence was as follows. The monkey self-initiated each trial by pressing a switch with the hand (TS). After a start period  of \myunits{400}{ms} a warning signal (WS) lighted up to focus the attention of the monkey. After another \myunits{400}{ms}, the cue (CUE-ON until CUE-OFF) informed the monkey either about the grip type (grip-first condition) or the force (force-first condition) required in this trial. The duration of cue presentation was \myunits{300}{ms}. It was followed by a preparatory delay period  of \myunits{1}{s}. The subsequent GO signal completed the information about force and grip, respectively, and in parallel asked the monkey to perform the movement by using the correct grip type and force to pull and then hold the object in a defined position window for \myunits{500}{ms}. Further periods: reaction time (RT) between the GO signal onset until the monkey released the switch (SR), movement time (MT) between switch release and object touch (OT), and pull time (PT) between OT and reaching the correct holding position. For correct trials the monkey was rewarded (RW) at the end of the holding period with a drop of apple sauce. See \figref{spectral-properties}A for a graphic representation of the task design.

\subsection*{Neuronal Recordings}
\label{sec:Methods-Neuronal-Recordings}
After completed training, a 100-electrode Utah array (Blackrock Microsystems, Salt Lake City, UT, USA) was chronically implanted in M1 and PM, contralateral to the working hand (for location see \figref{spectral-properties}C)  and overlapping rostral M1 and the posterior end of the dorsal premotor cortex (PMd) in monkeys L and T. The array of monkey N was placed more laterally covering the most medial part of the ventral premotor cortex (PMv). The \myunits{4 x 4}{mm} silicon based array consisted of 10-by-10 Iridium-Oxide electrodes, of which 96 were available for recording. The length of each electrode was \myunits{1.5}{mm}, with a \myunits{400}{\mu m} inter-electrode spacing. With this electrode length, we assume that the array enabled recording between the deep cortical layer III until the most superficial part of layer V. The distance between any pair of electrodes can be easily determined from the fixed geometric structure of the array. The surgery for array implantation was described in Riehle et al. (2013) and is briefly summarized below. The surgery was performed under deep general anesthesia using full aseptic procedures. A \myunits{30}{mm} x \myunits{20}{mm} craniotomy was performed over the motor cortex and the dura was incised and reflected. The array was inserted into the motor cortex between the central and arcuate sulci (Fig. 1C) using a pneumatic inserter (Blackrock Microsystems). It was then covered by a non-absorbable artificial dura (Preclude, Gore-tex). Ground and reference wires were inserted into the subdural space. The dura was then sutured back and covered with a piece of artificial absorbable dura (Seamdura, Codman). The bone flap was put back at its original position and secured to the skull by a titanium strip and titanium bone screws (Codman). The array connector was fixed to the skull on the hemisphere opposite to the implant. The skin was sutured back over the bone flap and around the connector. The monkey received a full course of antibiotics and analgesics after the surgery and recovered for one week before the first recordings.

Neuronal data were recorded using the 128-channel Cerebus acquisition system (NSP, Blackrock Microsystems). The signal from each active electrode (96 out of the 100 electrodes were connected) was preprocessed by a head stage (monkey L and T: CerePort plug to Samtec adaptor, monkey N: Patient cable, both Blackrock Microsystems) with unity gain and then amplified with a gain of 5000 using the Front End Amplifier (Blackrock Microsystems). The raw signal was obtained with 30 kHz time resolution in a range of \myunits{0.3}{Hz} to \myunits{7.5}{kHz}. From this raw signal, two filter settings allowed us to obtain on-line two different signals by using filters in two different frequency bands, the local field potential (LFP, low-pass filter at \myunits{250}{Hz}) and spiking activity (high-pass filter at \myunits{250}{Hz}). Here, only LFPs were analyzed, which were down-sampled at \myunits{1}{kHz}.

\subsection*{Power spectra}
\label{sec:Methods-Power-Coherence}
Power spectra were calculated using Welch's average periodogram algorithm using the \texttt{psd} function of the Python package \texttt{scipy}. We used windows of length $l=1024$ sample points (at \myunits{1}{kHz} sampling). Each window was tapered using a Hanning window. The time-resolved power spectra (spectrograms) were calculated using windows of length $l=512$ samples and an overlap of $500$ samples.

\subsection*{Definition of maps and vector fields}
\label{sec:Methods-Vector-Fields}
We calculated 5 different types of maps in order to visualize the spatial arrangement of oscillatory activity in the beta range on the array, and to provide a starting point for calculating multiple measures that characterize the arrangement. In a first step, we filtered the LFP signal on each electrode using a third-order Butterworth filter (pass band: \myunits{13-30}{Hz}) in a way that preserved the phase information (\texttt{filtfilt()} function of the Python package \texttt{scipy}). The filter setting was intentionally chosen broad such that it enabled us to identify the phase and the amplitude despite temporal variations of the beta oscillation amplitude and its center frequency. In order to compare the relative changes in amplitude between different electrodes, the amplitude of the LFP signal was then normalized across recording electrodes by computing the z-transform of the complete filtered LFP signal on an electrode-by-electrode basis.

In a next step, we calculated the instantaneous amplitude and phase of the normalized, filtered LFP time series $x_i(t)$ on each electrode $i$ by first constructing the analytic signal $X_i(t)=x_i(t)+j\ \mathcal{H}[x_i(t)]$, where $\mathcal{H}(\cdot)$ represents the Hilbert transform and $j^2=-1$. From $X_i(t)$, we obtained the instantaneous signal amplitude $a_i(t)$ by taking its modulo, and the instantaneous phase $\phi_i(t)$ by taking its argument (angle). We defined the maps $\ram{xy}=a_{i}(t)$ and $\rpm{xy}=\phi_{i}(t)$ by the instantaneous phases of the LFP, where $x\in\{0,\ldots,9\}$ and $y\in\{0,\ldots,9\}$ are the coordinates of the recording electrode $i$ of the Utah array in units of the inter-electrode distance of \myunits{400}{\mu m}.

In a further step we investigated whether, locally at each electrode and at each point in time, there is a spatially structured arrangement of the phases $\rpm{xy}$. To this end, in the remainder of this section, we defined three additional maps that we term the \textit{phase gradient} map $\pgm{xy}$, the \textit{directionality} map $\pdm{xy}$, and the \textit{gradient coherence} map $\gcm{xy}$ (cf.~\figref{methods-helper}A). The local spatial phase gradient at position electrodes $(x,y)$ was estimated based on a neighborhood $\bar\aleph_{xy}$ of its $k$-nearest neighbors in the same column $x$ or row $y$ (see \figref{methods-helper}B for a graphical representation). For border electrodes ($x\notin \{2,\ldots,7\}$ or $y\notin \{2,\ldots,7\}$), only existing electrodes were considered as nearest neighbors. In this manuscript we chose $k=2$ to obtain a smooth map of the local phase gradients. Let $\bar N_{xy}$ denote the cardinality of the set $\bar\aleph_{xy}$. We now constructed the phase gradient map as the average gradient $\textrm{d}|\phi(t)|/\textrm{d}x\cdot e^{j\alpha}$, between electrode $(x,y)$ and each of its neighbors $(i,j)$, where $\alpha$ denotes the angular direction between the electrode locations. The result is the map of \textit{phase gradients}
\begin{equation}
 \pgm{xy}=
 \bar N_{xy}^{-1}\sum_{(i,j)\in 
\bar\aleph_{xy}}\frac{\rpm{ij}-\rpm{xy}}{\sqrt{(i-x)^2+(j-y)^2}}\cdot 
e^{j\alpha_{ij}}
 \approx \nabla \rpm{xy}.
 \label{eq:pgm-def}
\end{equation}
Based on the average frequency of the beta oscillation $f_\beta$, we can easily derive the phase velocity field $\pvm{xy}=2\pi f_\beta\ |\pgm{xy}|^{-1}$, which indicates the phase velocity of a planar wave front running through the point $(x,y)$. Here, \myunits{f_\beta=21.5}{Hz} was chosen as the mean frequency of the respective beta bands of the monkeys (see above). An estimate of the macroscopic phase velocity can be obtained by calculating the average $\velocity=\overline{|\pvm{xy}|}$ across electrodes. Next, we defined the phase directionality map
\begin{equation}
 \pdm{xy}=|\pgm{xy}|^{-1}\pgm{xy}
 \label{eq:pdm-def}
\end{equation}
by normalizing the the vectors of the phase gradient map $\pgm{xy}$ to unit length. It indicates only the direction of the local phase gradient, independent of its magnitude. Finally, we defined the gradient coherence map as an average of the directionality map in a neighborhood $\hat\aleph_{xy}$ of all $k$-nearest neighbors of cardinality $\hat N_{xy}$ (cf.~\figref{methods-helper}B):
\begin{equation}
 \gcm{xy}=\hat N^{-1}\sum_{i,j\in \hat\aleph_{xy}}\pdm{ij}.
 \label{eq:gcm-def}
\end{equation}
It represents a second order measure of the gradient field and serves two purposes. The direction of each entry in $\gcm{xy}$ provides a smoothed version of the vector field $\pgm{xy}$, which is better suited for visualization due to the rather sparse sampling of activity. More importantly, the magnitude of the vectors in $\gcm{xy}$ indicate whether, locally, phase gradients point in the same direction (independent of the magnitudes of the gradients).
\subsection*{Quantification of observed phase patterns}
\label{sec:Methods-Wave-Pattern-Quantification}
Based on the phase map $\rpm{xy}$ and the three vector fields $\pgm{xy}$, $\pdm{xy}$, and $\gcm{xy}$, we now introduced 6 measures that quantitatively describe the spatial arrangement of phases on the array at each time point. These measures will later on serve as a basis to classify the phase pattern, i.e.\ the spatial arrangement of phases in $\rpm{xy}$, in an automatized manner. In the following, let $\aleph$ denote the set of all used electrodes in a given recording, and $N=|\aleph|$ its cardinality.

\paragraph{Circular variance of phases.}
One phase pattern commonly observed is the one where all electrodes are fully synchronized at near-zero phase lag. Therefore, we introduced the \textit{circular variance of phases}
\begin{equation}
 \cvp=N^{-1}\sum_{(i,j)\in \aleph}e^{j\rpm{ij}}\in [0,1]
 \label{eq:cvp-def}
\end{equation}
as a measure to determine the similarity of the phase across the array. Here, $\cvp=0$ indicates that an identical phase $\rpm{xy}$ observed at each electrode, whereas $\cvp=1$ indicates that phases are uniformly distributed across the array.

\paragraph{Circular variance of phase directionality.}
In order to measure the degree to which phase gradients are globally aligned across the grid, we introduced the \textit{circular variance of the phase directionality}
\begin{equation}
 \cvg=N^{-1}\sum_{(i,j)\in \aleph}\pdm{ij}\in [0,1].
 \label{eq:cvg-def}
\end{equation}
A perfect planar wave is observed if $\cvg=0$, i.e., all phase gradients point in the same direction (independent of the magnitude of the gradients). This measure is similar to the PGD measure defined by \citet{Rubino06_1549}.

\paragraph{Local gradient coherence.}
In order to determine whether locally (within $\hat\aleph_{xy}$) phase gradients point in a particular direction, we considered the average length of the vectors forming the gradient coherence vector field $\gcm{xy}$ and defined the \textit{local gradient coherence}
\begin{equation}
 \mlc=\left|N^{-1}\sum_{(i,j)\in \aleph}|\gcm{ij}|^{-1}\gcm{ij}\right|\in [0,1].
 \label{eq:mlc-def}
\end{equation}
It has a value of $1$ if in each neighborhood all phase gradients are perfectly aligned. In particular, we note that $\cvg=1\Rightarrow\mlc=1$.

\paragraph{Gradient continuity.}
Along the same argument we may ask even more strictly whether phase gradients locally not only point in a similar direction, but whether in fact they tend to form continuous lines. To this extend we defined in the following the \textit{gradient continuity} $\cnt$. For each point $(x,y)$, we determined the direction of the phase gradient from $\pdm{xy}$ and calculated the electrode at position $(i,j)$ that the gradient points to. We then calculated the scalar product $s_{xy}=\langle\pdm{xy},\pdm{ij}\rangle$, which is 1 if the two lines form a continuous line, and $0$ if the gradients are orthogonal to each other. When averaging across all electrodes we obtain
\begin{equation}
 \cnt=N^{-1}\sum_{(i,j)\in \aleph}s_{xy}\in [-1,1].
 \label{eq:cnt-def}
\end{equation}

\paragraph{Circular arrangements.}
Besides spatial phase patterns that appear either homogeneous or linear, we observed phase patterns that appear radial or circular with respect to the center of the array. To capture such spatial arrangements, we defined the two measures \textit{radial-parallel alignment} $\prl$ and \textit{radial-orthogonal alignment} $\ort$. We first constructed a vector $l_{xy}$ normalized to length $|l_{xy}|=1$ pointing from the center of the array (i.e., at an imaginary electrode located at electrode coordinates $(4.5,4.5)$) to each electrode at position $(x,y)$. We then defined the scalar product $s^\parallel_{xy}$ between the phase gradient $\pdm{xy}$ at $(x,y)$ and $l_{xy}$, and the scalar product $s^\perp_{xy}$ between $\pdm{xy}$ and the perpendicular direction $l_{xy}e^{j\pi/2}$. Then
 \begin{align}
 \prl&=N^{-1}\sum_{(i,j)\in \bar\aleph}\lvert s^\parallel_{xy}\rvert\in [0,1],\ \ \ \textrm{and}\\
 \ort&=N^{-1}\sum_{(i,j)\in \bar\aleph}\lvert s^\perp_{xy}\rvert\in [0,1].
 \label{eq:prl-ort-def}
 \end{align}
$\prl=1$ indicates that phase gradients point inward or outward, as observed for a radial phase pattern, whereas $\ort=1$ indicates that phase gradients point in the orthogonal direction, as observed for a circular phase pattern.

\subsection*{Classification of spatial phase patterns}
\label{sec:Methods-Classification}
Equipped with the 6 measures defined above, we were now able to classify the spatial pattern observed at each point in time according to the salient patterns that were visually observed. The phase patterns we distinguished are termed the \textit{planar wave}, \textit{synchronized}, \textit{random}, \textit{circular}, and \textit{radial} patterns. A detailed qualitative description of these patterns is presented in the \nameref{sec:Results}.

%
%
\begin{figure}
 \centering
  \begin{minipage}[t]{0.63\textwidth}
    \ \\
    \includegraphics[width=\textwidth]{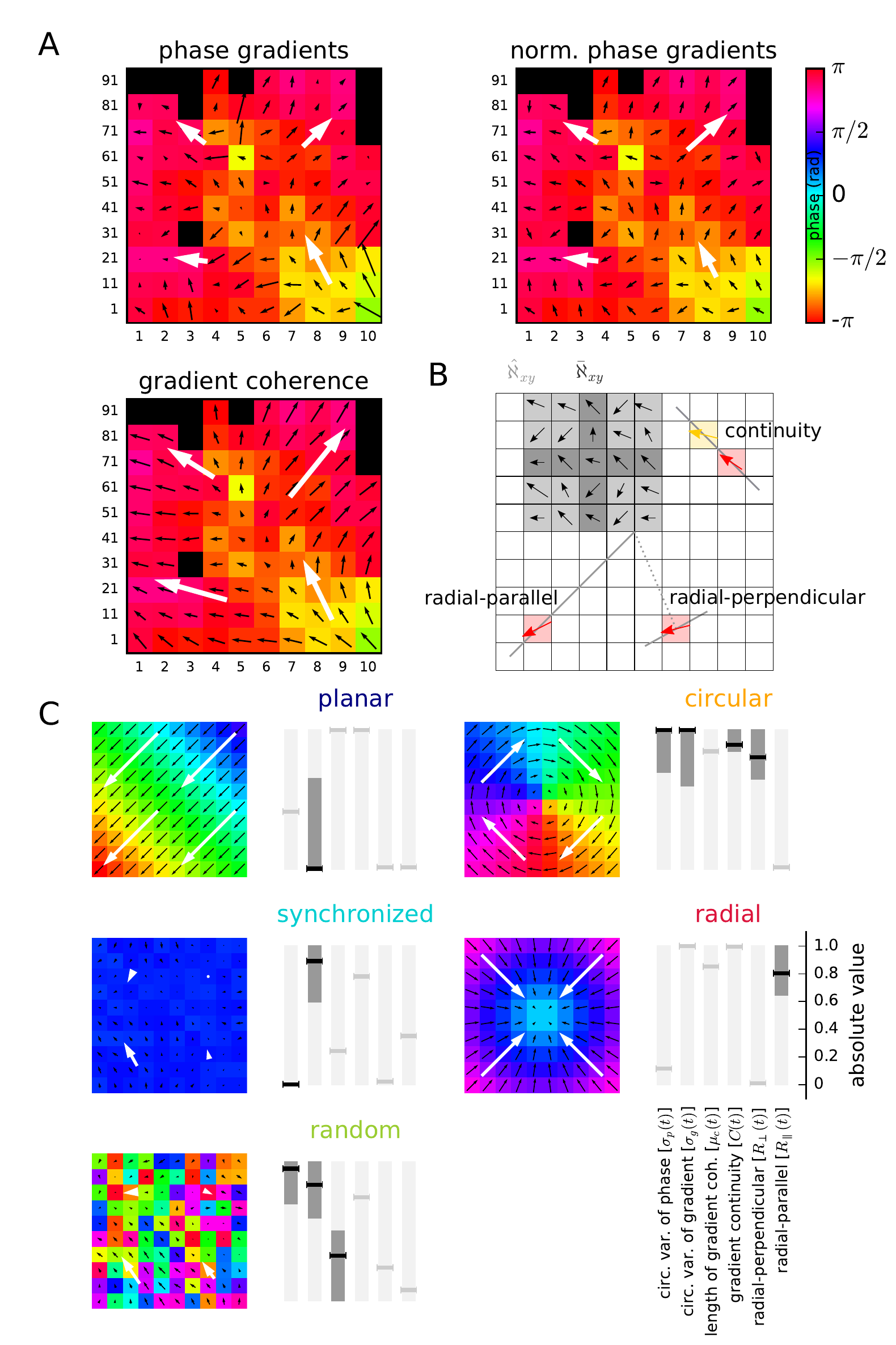}
  \end{minipage}\hfill
  \begin{minipage}[t]{0.34\textwidth}
    \caption{\textbf{Explanation of the methodology.} \textbf{A.} Example phase map $\rpm{xy}$ (colors) of monkey L (session identifier l101013-002) with overlay of phase gradient map $\pgm{xy}$ (top left), phase direction map $\pdm{xy}$ (top right) and gradient coherence map $\gcm{xy}$ (bottom left) shown as black arrows per electrode. White large arrows: corresponding quadrant-averaged maps. \textbf{B.} Sketch of the 10x10 electrode array (squares) to illustrate the neighborhoods $\hat\aleph_{xy}$ (light gray) and $\bar\aleph_{xy}$ (dark gray), and the calculation of the gradient continuity $\cnt$ (top right), radial-parallel alignment $\prl$ (bottom left), and radial-orthogonal alignment $\ort$ (bottom right). $\cnt$: angle between the gradient (red arrow) at a chosen electrode (red square) and the gradient at the square that the red gradient points to (yellow arrow and square). $\prl$ and $\ort$: angle between the gradient (red arrow) at a chosen electrode (red square) and the corresponding solid gray line (parallel to and perpendicular to a line through the array center and the electrode, respectively). \textbf{C.} Calculated values for the 6 measures (represented by horizontal line markers; from left to right: $\cvp$, $\cvg$, $\mlc$, $\cnt$, $\prl$, and $\ort$) for artificial data modeling ideal realizations of the 5 phase patterns (planar wave, synchronized, random, circular, and radial). A small amount of noise was added to the synchronized phase pattern to avoid division by zero in calculating measures. Black markers: measure was used to detect the corresponding phase pattern. Dark areas indicate the range of valid values for that measure required for a positive classification according to the threshold criteria  (cf.,~\tabref{Class-Thresholds}). Light gray markers: measure was not used to detect the corresponding pattern. Region of valid values for measure $\cvp$ for synchronized patterns is close to 0.} \label{fig:methods-helper}
  \end{minipage}\hfill
\end{figure}

To disambiguate the 5 phase patterns we manually define specific thresholds $\theta_1-\theta_8$ on those measures that capture relevant characteristics of the patterns. Specifically, $\theta_1$ and $\theta_2$ are thresholds on $\cvp$, $\theta_3$ and $\theta_4$ on $\cvg$, $\theta_5$ on $\mlc$, $\theta_6$ on $\cnt$, $\theta_7$ on $\ort$, and $\theta_8$ on $\prl$. The threshold conditions for each phase pattern are summarized in \tabref{Class-Thresholds}. In the following we summarize the rationale for choosing these threshold conditions. \textit{Planar wave} patterns are characterized by a non-zero phase-gradient that points in the same direction at each electrode, and are thus well characterized by a small value of $\cvg$. Perfectly \textit{synchronized} patterns exhibit the same phase at each electrode (small $\cvp$), and the grid-averaged phase gradient is random (large $\cvp$). \textit{Random} patterns, on the other hand, show a random phase at each electrode (large $\cvp$), and also the phase gradient is random not only when considering the complete grid (large $\cvg$), but also locally in each neighborhood $\hat\aleph_{xy}$ of an electrode (large $\mlc$). \textit{Circular} patterns share the feature of displaying all phases and phase gradients across the array with the random pattern (large $\cvp$ and $\cvg$), however, they exhibit a high continuity of the gradient coherence field (high $\cnt$) and phase gradients are arranged in a circular fashion around the grid center (high $\ort$). Similarly, \textit{radial} patterns exhibit phase gradients that are arranged in an outward or inward pointing direction with respect to the grid center (high $\prl$), however, in contrast to the circular pattern the distributions of phases and phase gradients, and the gradient continuity are less characteristic of this pattern, and therefore are not used. Due to the low spatial sampling, in rare cases a pattern could fulfill two conditions, e.g., a circular pattern could sometimes also be viewed as random pattern. In such cases, we defined an order of administering the tests of the patterns (from first to last: planar, radial, synchronized, circular and random patterns) and assigned the pattern to the first successful classification in that order. In this study the following values were used: $\theta_1=0.15$, $\theta_2=0.7$, $\theta_3=0.5$, $\theta_4=0.6$, $\theta_5=0.5$, $\theta_6=0.85$, $\theta_7=0.65$, and $\theta_8=0.65$.

{%
\begin{center}
\begin{table}
\caption{\textbf{Thresholds used to classify phase patterns.} A time point was classified as one of the 5 listed phase patterns if all corresponding threshold criteria were met. Tests were administered in the order given by the table, and the first match is chosen as the classification.}
\ \\
\label{tab:Class-Thresholds}
\begin{tabular}{|l|l|}\hline
\textbf{phase pattern} & \textbf{thresholds}\\\hline\hline
planar wave & $\cvg<\theta_3$\\\hline
radial & $|\prl|>\theta_{8}$\\\hline
synchronized & $\cvp<\theta_1$, and $\cvg\geq\theta_4$\\\hline
circular & $\cvp\geq\theta_2$, $\cvg\geq\theta_4$, \\
\ & $\cnt\geq\theta_6$ and $|\ort|\geq\theta_7$\\\hline
random & $\cvp\geq\theta_2$, $\cvg\geq\theta_4$, \\
\ & and $\mlc\leq\theta_5$\\\hline
\end{tabular}
\end{table}
\end{center}
}

\end{methods}

\paragraph{Acknowledgements} This work was supported by the Helmholtz Portfolio Supercomputing and Modeling for the Human Brain (SMHB), EU grants 604102 and 720270 (Human Brain Project, HBP), EU Grant 269912 (BrainScaleS), Deutsche Forschungsgemeinschaft Grants DE 2175/2-1 and 1753/4-2 Priority Program (SPP 1665), ANR-GRASP, Neuro\_IC2010, CNRS-PEPS, and the Riken-CNRS Research Agreement.\\

%
%
\references{waves.bib}   

%
%

%
%
\newpage
\section*{Supplemental Information}
\label{sec:Supplement}
\pdfbookmark[0]{Supplemental Information}{sec:Supplement-pdf}

\makeatletter
\setcounter{figure}{1}
\renewcommand\thefigure{S\arabic{figure}}
\makeatother

%
%
\begin{figure}[h]
 \centering
 \includegraphics[width=\linewidth]{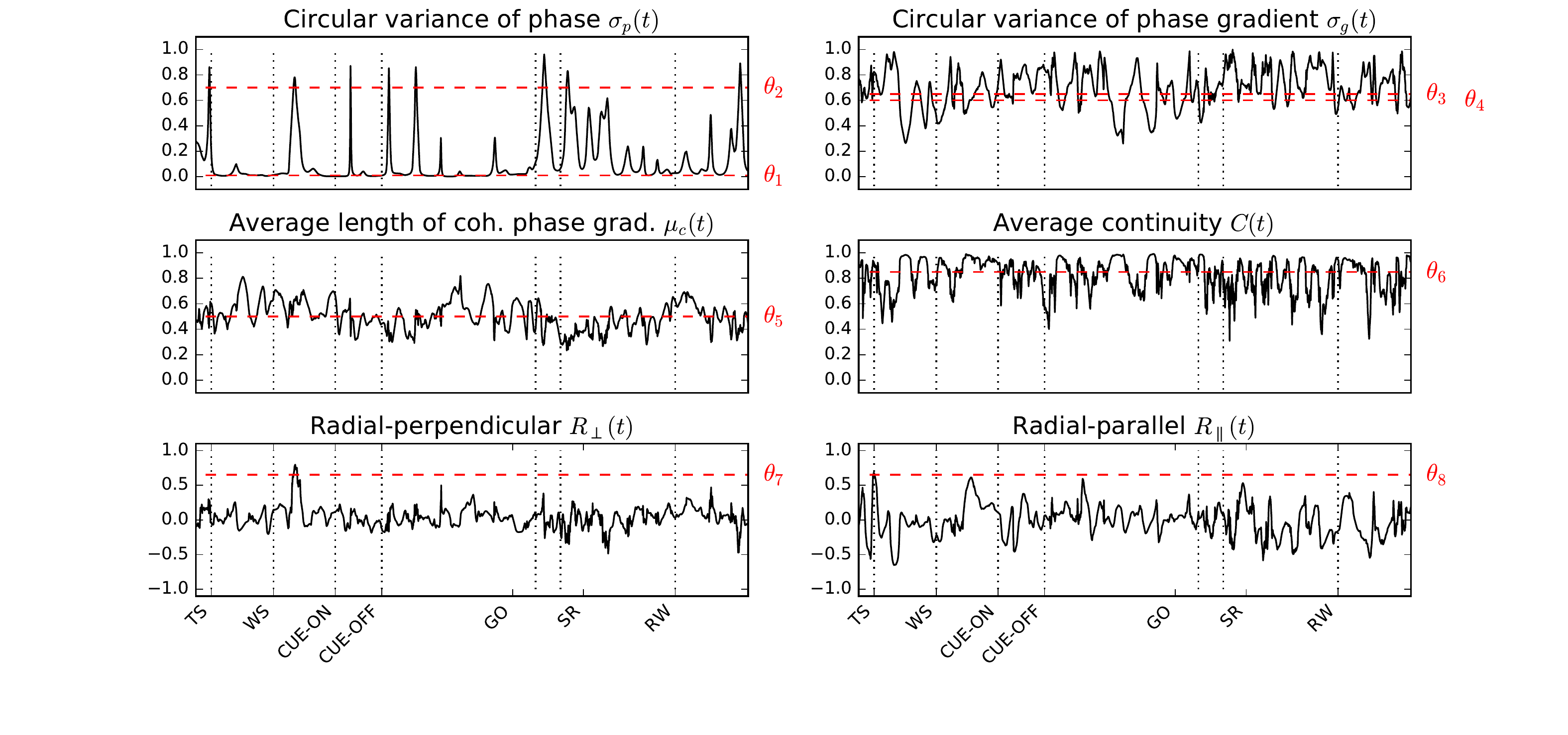}
 \caption{\textbf{Thresholds used for automatic classification of the example trial.}  Graphs of the time series (corresponding to the trial shown in panel \figref{phase-patterns}A) of the 6 measures used for detection of the different patterns (top to bottom: $\cvp$, $\cvg$, $\mlc$, $\cnt$, $\prl$, and $\ort$). Red dashed lines indicate the thresholds $\theta_1$ to $\theta_8$ used to detect and classify the phase patterns at each time point.} \label{fig:phase-patterns-thresholds}
\end{figure}

%
%
\begin{figure}[h]
 \centering
 \includegraphics[width=\linewidth]{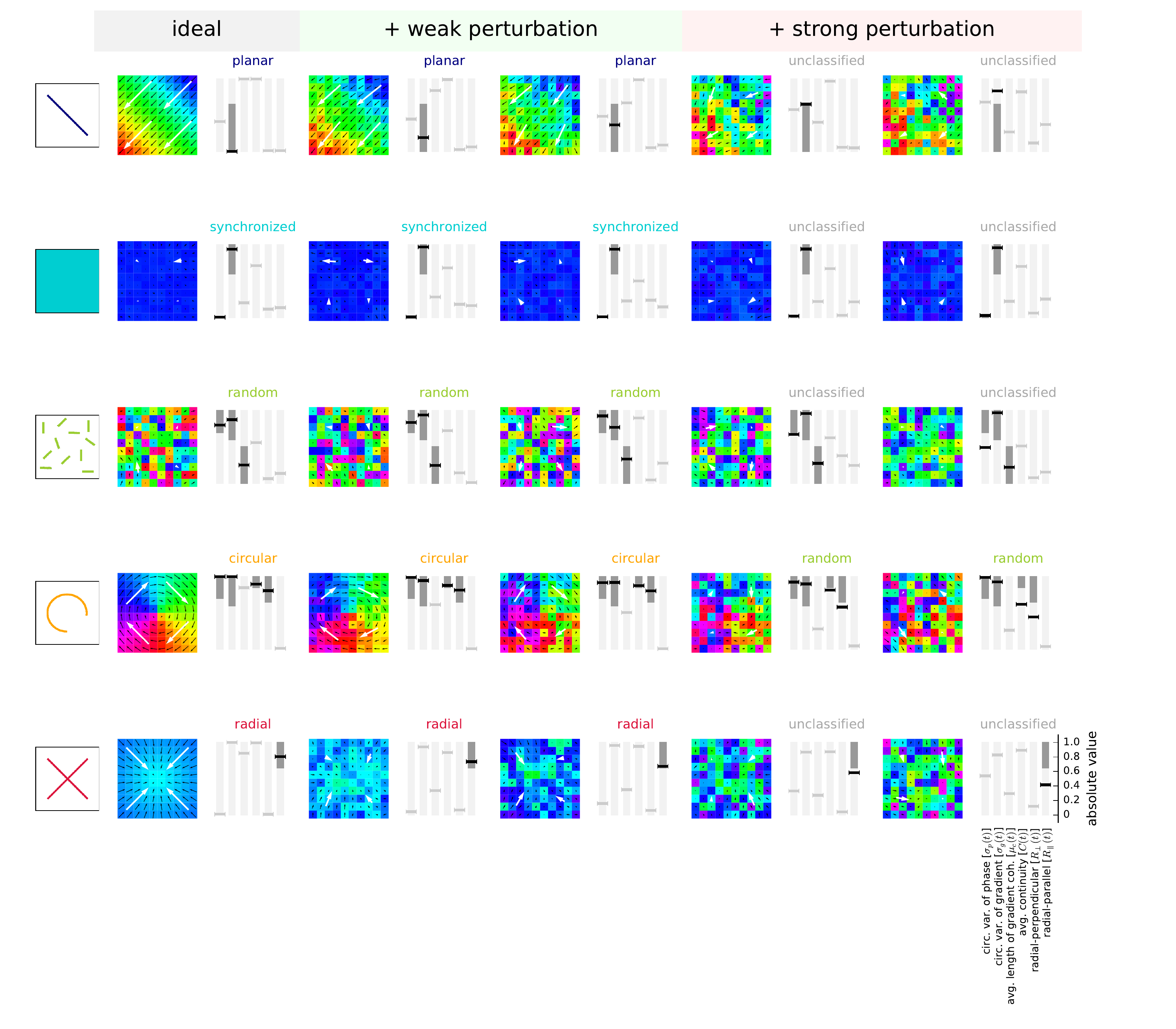}
 \caption{\textbf{Visual illustration and justification of thresholds used for automatic classification.} For each of the idealized planar wave, synchronized, random, circular, and radial patterns (rows) 5 artificially generated realizations are shown (cf.,~\figref{methods-helper}): one ideal pattern (left column), two patterns with weak perturbation (weak additive noise) which we considered as valid representatives of that pattern type (second and third column), and two patterns with strong perturbation (strong additive noise) which we did not consider as valid representatives of that pattern type (fourth and fifth column). Absolute values of the 6 measures are shown to the right of each pattern (represented by horizontal bar markers; from left to right: $\cvp$, $\cvg$, $\mlc$, $\cnt$, $\prl$, and $\ort$). Black markers: measure was used to detect the corresponding phase pattern. Dark areas indicate the range of valid values for that measure required for a positive classification according to the threshold criteria. Light gray markers: measure was not used to detect the pattern treated in the corresponding row. Text above panels: result of automatic pattern classification. Due to the specific choice of destroying patterns by adding noise (as opposed to, e.g., randomizing phase gradients), not every measure is necessarily affected by increased noise levels. Exceptions: (i) Ideal synchronized pattern (row 2, column 1) has minimal additional noise to avoid division by zero when calculating measures, (ii) for random patterns (row 3), the amount of additive noise is decreased from left to right to drive the state from fully randomized phases towards the synchronized state.} \label{fig:threshold-setting}
\end{figure}

%
%
\begin{figure}[h]
 \centering
 \includegraphics[width=\textwidth]{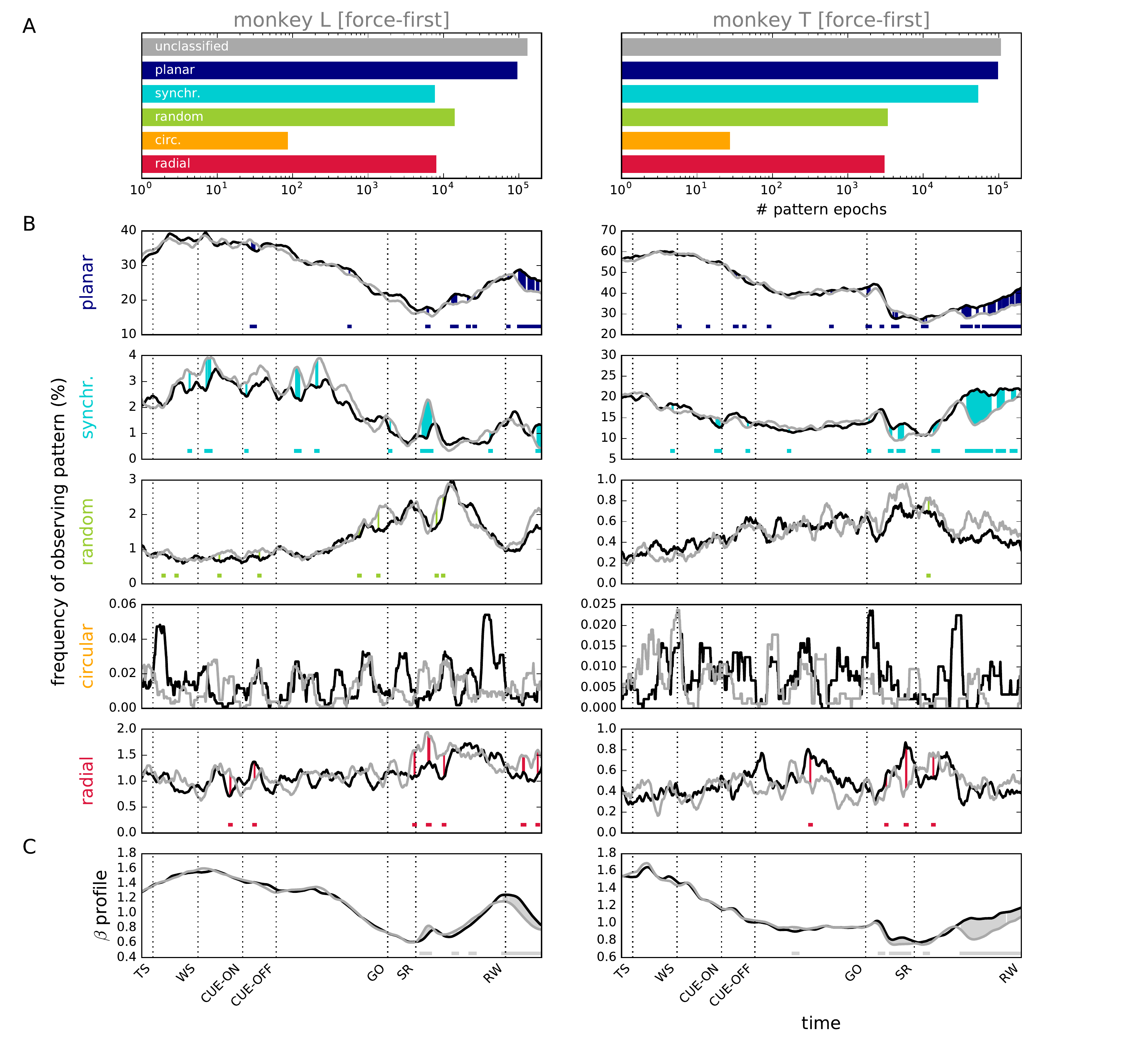}
 \caption{\textbf{Behavioral correlates and relation to average beta amplitude for the force-first condition.} \textbf{A.} Number of epochs of a phase pattern detected in at least 5 consecutive time frames, i.e., \myunits{5}{ms} (bars from top to bottom: unclassified, planar wave, synchronized, random, circular, radial pattern) for monkey L (left) and T (right). Data were obtained from all selected recording sessions including inter-trial intervals. \textbf{B.} Time-resolved probability of observing a specific phase pattern (rows) during the trial. Statistics were computed across all force-first trials of all recording sessions for both monkeys ($N=15$) and smoothed with a box-car kernel of length $l=$\myunits{100}{ms}. Trials were separated into side-grip (SG) trials (black) and precision-grip (PG) trials (gray). In contrast to \figref{behavioral-correlates}, differences between these trial types occur only after the grip information presented at GO. Color shading between curves and colored bars indicate time periods where SG and PG curves different significantly (Fisher's exact test under the null hypothesis that, for any time point, the likelihood to observe a given phase pattern is independent of the trial type, $p<0.05$). \textbf{C.} Beta amplitude profile (envelope) pooled across all SG (black) and PG (gray) trials (same data as in panel B). The amplitude profile $\amplitude$ of a single trial is calculated as the time-resolved instantaneous amplitude $\ram{xy}$ of the beta-filtered LFP averaged across all electrodes $(x,y)$, and measures the instantaneous power of the beta oscillation in that trial. Gray shading between curves and horizontal bars indicate time periods where SG and PG curves differ significantly (t-test under the null hypothesis that the distributions of electrode-averaged single trial amplitudes $\ram{xy}$ at each time point $t$ are identical for SG and PG trials, respectively, $p<0.05$).}
 \label{fig:behavioral-correlates-force-first}
\end{figure}
\ \\

\end{document}